\documentclass[
 reprint,
 amsmath,amssymb,
 aps,
]{revtex4-2}

\usepackage{xcolor}
\usepackage{graphicx}
\usepackage{dcolumn}
\usepackage{bm}
\usepackage{hyperref}
\usepackage[mathlines]{lineno}

\DeclareMathOperator{\diag}{diag}

\usepackage{slashed} 
\usepackage{bm} 
\usepackage{breqn}
\usepackage{mathtools}
\usepackage{multirow} 
\usepackage{indentfirst} 
\usepackage{lipsum} 
\usepackage{mathrsfs} 
\usepackage{float} 
\usepackage{subfigure}
\usepackage{csquotes} 
\usepackage{enumitem} 

\usepackage{comment} 
\begin{document}

\preprint{APS/123-QED}

\title{Bayesian analysis of the properties of hybrid stars with the Nambu--Jona-Lasinio model}

\author{Antoine Pfaff}
\email{a.pfaff@ip2i.in2p3.fr}

\author{Hubert Hansen}
\affiliation{Univ. Lyon\char`,{} Univ. Claude Bernard Lyon 1\char`,{}
CNRS/IN2P3\char`,{} Institut de Physique des 2 Infinis de Lyon\char`,{} UMR 5822\char`,{} 69622 Villeurbanne\char`,{} France
} 
    
\author{Francesca Gulminelli}

\affiliation{
LPC Caen IN2P3-CNRS/EnsiCaen and Universite Caen France
}%



\date{\today}

\begin{abstract}

The possible presence of deconfined matter in the cores of massive neutron stars is the subject of a large debate. In this context, it is important to set limits on the size and characteristics of a hypothetical quark core compatible with the present astrophysical constraints.

To this aim, we present a Bayesian analysis of the properties of non-rotating hybrid stars at equilibrium with quark matter cores, as described by the SU(3) Nambu--Jona-Lasinio (NJL) model.
The hadronic phase is described by a unified metamodeling approach, with a prior parameter space covering the present uncertainties on nuclear matter properties with nucleonic degrees of freedom. 
The parameter space of the NJL model includes vector-isoscalar and vector-isovector couplings and, additionally, an effective bag constant for the quark pressure is introduced as a free parameter.
The phase transition is assumed to be first order with charge neutral phases, following the Maxwell construction. Our Bayesian framework includes filters on the experimental and theoretical low-density nuclear physics knowledge (atomic masses, ab initio calculations of the EoS) and high density  constraints from astrophysical observations (maximum mass of J0348+0432, binary tidal deformability of the GW170817 event).
We find that microscopic vector interactions play an important role in quark matter in order to stiffen the equation of state sufficiently to reach high star masses, in agreement with previous studies. Even within a very large prior for both the hadronic phase and the quark phase and the important freedom brought by the effective bag constant, our posterior quark cores tend to be relatively small and only appear in very heavy stars ($M\gtrsim 2 M_\odot$). 
Coincidentally, the inclusion of the nucleon-quark transition (deconfinement transition) only weakly affects the radii of compact stars, foreshadowing a very low observability of a possible phase transition using x-ray radii measurements.
\end{abstract}
\maketitle

\section{\label{Introduction}Introduction}

The color confinement and the characterization of the phase structure of quantum chromodynamics (QCD) at finite temperature and density remains a challenge to this day, after years of experimental and theoretical efforts.
If not at all well understood theoretically, the color deconfinement of QCD is at least quite well described at finite temperature and zero or small density, for example by lattice QCD (LQCD) calculations or in earth bound experiments in hadronic colliders. 
It occurs as a crossover deconfinement phase transition (PT) around a temperature  $k_B T_0\approx160$ MeV \cite{WB2006,WB2006bis,WB2010,HotQCD2012,HotQCD2014}.

One of the most decisive results in QCD would be to determine its degrees of freedom and equation of state (EoS) at finite temperature and large density, a region where LQCD is not applicable in general and experiments cannot easily reach as of yet. 
Hence, the proof of the existence and the characterization of a deconfinement transition in compact stars would be a fundamental achievement.

It is often assumed that neutron stars may be a laboratory to study high density quark phases of QCD, albeit a very distant one, despite the fact that no direct probe of deconfined matter can be measured (in contrast, probes such as the dileptons can give us more direct information in colliders). 
One has to rely on less direct information such as x-ray spectrum or Gravitational Wave (GW) signals during mergers that need to be interpreted via effective models of QCD to gain information on the QCD transition characteristics; see Refs. \cite{Miao2020,Tang2021,Li2021,Xie2021,Blaschke2021} for some recent works.

In the absence of a satisfactory and controlled computation
of the EoS from first-principle QCD with quark and gluonic degrees of freedom~\cite{Fu2020, Otto2020}, a large number of phenomenological approaches have been proposed, such as bag models, the quark-meson coupling model, and Nambu--Jona-Lasinio (NJL) models~\cite{Buballa,Pereira2016,Li2018,Zacchi2019,Isserstedt2019,Motornenko2020,Morimoto2020,Tan2020,Alaverdyan2021,Lopes2021,Blaschke2017,Cierniak2018}, all dependent on different and largely unconstrained parameters. To appreciate the model dependence of the predictions, it is therefore very important to perform studies where the parameter space is largely studied, at least in the framework of a specific effective
approach~\cite{Biguet,Cierniak2021,Ferreira2021}. In this regard, the Bayesian tool has proved to be essential in order to extract useful information on the parameter space from the available experimental constraints \cite{Millerbayesian,Blaschke2015,AlvarezCastillo2020,Alvarez2016,Ayriyan2019,Blaschke2020}.

In this paper, we propose an approach along those lines within a Bayesian framework based on hybrid EoSs with the Nambu--Jona-Lasinio (NJL) model on the quark matter side and a general metamodelization of nucleonic matter on the hadronic side. 
We confirm previous results with similar techniques and models \cite{Ferreira2020,Ferreira2021} showing that it is possible to have quark cores in high-mass neutron stars respecting the current astrophysical constraints, but that unfortunately the observations are still not accurate enough to decide on the existence of a deconfined core. These conclusions are reinforced by the fact that, with respect to the work of Ref.\cite{Ferreira2020,Ferreira2021}, we use a fully controlled nucleonic EoS, that contains a unified treatment of the NS crust, and includes both constraints from nuclear mass measurements and from ab-initio nuclear theory.

\medskip
This paper is organized as follows. 
In Sec. \ref{sec::Method}, we will review our methods: the models used on the hadronic and quark sides, the modelization of the transition as first order, particularly focusing on the role of the so-called bag constant, 
and finally the Bayesian framework.
In Sec. \ref{sec::results}, we will present our results: the posterior distributions of the parameters and the resulting EoS are first discussed, then we will take a look at the parameters and observables correlations, finally concluding on the possibility of having quark cores in heavy neutron stars.

\section{Method} \label{sec::Method}

In this section, we present the models used to construct the hybrid equations of state, as well as the method employed to implement the subsequent Bayesian analysis of hybrid stars properties. Throughout all this paper, we use the natural unit system in which $\hbar=c=1$.

\subsection{The nuclear meta-model} \label{sec:meta}

In order to explore consistently the current theoretical uncertainty on the nuclear equation of state, we use at low densities a meta-model of nuclear matter~\cite{Margueron_2018,Jerome_2018} extended to include surface terms~\cite{Carreau_2019,Hoa1,Hoa2} to allow a consistent and unified description of the crust. This approach is based on the hypothesis that the total baryonic density $n_B$ can be simply written as the sum of the neutron $n_n$ and proton $n_p$ densities, ${n_B=n_p+n_n}$. The energy per particle of infinite nuclear matter $e_N$ is expressed as a Taylor expansion in the rescaled baryon density ${x=(n_B-n_{sat})/3n_{sat}}$ and the isospin asymmetry ${\delta = (n_n-n_p)/n_B}$ around the saturation density of symmetric nuclear matter $n_{sat}$:

\begin{eqnarray}\label{bulk}
e_N(n_B,\delta)&&\,=E_{sat}+\frac{1}{2!}K_{sat}x^2+\frac{1}{3!}Q_{sat}x^3+\frac{1}{4!}Z_{sat}x^4 \nonumber \\
&& +\,\delta^2\Big(E_{sym}+L_{sym}x+\frac{1}{2!}K_{sym}x^2 \nonumber \\
&& \qquad\quad+\frac{1}{3!}Q_{sym}x^3+\frac{1}{4!}Z_{sym}x^4\Big) + ...
\end{eqnarray}
The coefficients of this expansion, also known as the nuclear empirical parameters (NEPs), entirely determine the nuclear density functional as long as it remains an analytic function, which is the case as long as no phase transition occurs and the EoS remains nucleonic in nature.

To improve the convergence of the expansion, a non-relativistic kinetic term (which involves non integer powers of $x$ and $\delta$) is added, and an exponential correction factor governed by an additional parameter~$b$ is applied in order to control the behavior of the EoS in the zero-density limit (see~\cite{Carreau_2019,Antic2019} for details). 
The final expression for the energy per baryon reads:
\begin{eqnarray}
e_N&=&  \frac{3}{20m}\left(\frac{3\pi^{2}n_B}{2}\right)^{2/3}
\bigg[ \left( 1+\kappa_{sat}\frac{n_B}{n_{sat}} \right) f_1
+ \kappa_{sym}\frac{n_B}{n_{sat}}f_2\bigg] \nonumber  \\
&+&\sum_{\alpha=0}^4 ( V_{\alpha}^{is}+ V_{\alpha}^{iv} \delta^2) \frac{x^\alpha}{\alpha!}
-(a^{is}+a^{iv}\delta^2) x^{5} e^{ -b\frac{n_B}{n_{sat}}},
\label{eq:vELFc}
\end{eqnarray}
%
where $m=939$~MeV is the nucleon mass, and the isospin dependence of the kinetic energy term is governed by the functions: 
\begin{eqnarray}
f_1(\delta) &=& (1+\delta)^{5/3}+(1-\delta)^{5/3} \, \\
f_2(\delta) &=& \delta \left( (1+\delta)^{5/3}-(1-\delta)^{5/3} \right) .
\end{eqnarray}

The interaction parameters $V_\alpha^{is}$ and $V_\alpha^{iv}$ 
of the meta-functional Eq.(\ref{eq:vELFc}) are one-to-one related to each of the NEPs in Eq.(\ref{bulk}), while the two parameters  $\kappa_{sat}$ and  $\kappa_{sym}$ are linked to the Landau effective mass  $m^\star_{sat}$ and isospin mass splitting $\Delta m^\star_{sat}$ at saturation. The parameters $a^{is}$ and $a^{iv}$ are entirely fixed by the condition at zero density. Explicit expressions for each of these quantities can be found in Ref.\cite{Margueron_2018}.
We will gather the full parameter set in a compact form with the quantity~\textbf{X}$_N$:
\begin{eqnarray}
\textbf{X}_N 
= \{n_{sat},&&E_{sat},K_{sat},Q_{sat},Z_{sat},E_{sym},L_{sym}, \\
&&K_{sym},Q_{sym},Z_{sym},m^\star_{sat},\Delta m^\star_{sat},b\} \nonumber
\end{eqnarray}

In order to correctly describe the inhomogeneous structure corresponding to the crust of neutron stars, it is crucial to take into account the finite size effects such as surface and Coulomb, associated with the crust energetics in addition to the bulk part governed by Eq.(\ref{bulk}). 
We use a compressible liquid drop model~(CLDM) to describe the mass of a finite nucleus with $Z$ protons and ${A=N+Z}$~nucleons as \cite{Carreau_2019,Hoa1}
%
\begin{equation}
M(A,Z)=N m_n+Z m_p +e_N(n_{0},I)+E_c+E_s, \label{eq:enuc}
\end{equation}
%
%
where $m_{n(p)}$ are the bare neutron and proton masses,  $n_{0}$~is the equilibrium bulk density of the nucleus which, in the vacuum, corresponds to the equilibrium density of nuclear matter at asymmetry ${\delta=I=(N-Z)/A}$, ${E_c = 3e^2Z^2/5R}$ is the Coulomb energy of a uniformly charged sphere of radius ${R=(3 A/(4\pi n_0))^{1/3}}$, and the surface energy $E_s$ is given by
\begin{equation}
E_s(A,Z) = 4\pi R^2\sigma + 8\pi R \sigma_c.
\end{equation}
The isospin dependent surface and curvature tensions~$\sigma,\sigma_c$ are taken as \cite{Ravenhall1983,LattimerSwesty,Carreau_2019,Hoa1}: 
\begin{eqnarray}
\sigma(A,Z) &=& \sigma_0\frac{2^{4}+b_s}{(Z/A)^{-3}+b_s+(N/A)^{-3}},  \label{eq:sigma}\\
\sigma_c(A,Z) &=& 5.5\, \sigma_{0,c} \frac{\sigma}{\sigma_0} \left(\beta -\frac{Z}{A}\right) \label{eq:sigmac},
\end{eqnarray}
The additional parameters  ${\textbf{X}_\sigma=\{\sigma_0,\sigma_{0,c},b_s,\beta\}}$ entering {Eqs.(\ref{eq:sigma}) and (\ref{eq:sigmac})} should in principle be added to our parameter set in the nucleonic sector.
However, since the mass of nuclei is inherently dependent on both bulk and surface effects, our precise experimental knowledge of nuclear masses brings strong correlations between the parameters describing the two effects.
We take advantage of this correlation to fix the   surface parameters \textbf{X}$_\sigma$  for each parameter set \textbf{X}$_N$, by performing a fit of the experimental masses \cite{AME2012} using a least-squares method.

It was verified in~\cite{Margueron_2018} that an expansion up to fourth order in $x$ and first order in $\delta^2$, such as the one proposed by Eq.(\ref{eq:vELFc}), has enough flexibility to accurately reproduce the behavior of a very large class of nuclear models up to densities a few times saturation density. Therefore, this meta-model can be used to freely explore the theoretical uncertainty on the nuclear EoS, allowing an in-depth study of the role of each NEP independently of the correlations that might emerge within a specific choice of the nuclear energy functional. In this spirit, and taking into account the current experimental knowledge on the lowest order NEP, we consider a flat prior distribution for the parameters in \textbf{X}$_N$ within intervals given in Table~\ref{nucprior}. 

\begin{table*}[htbp]
\centering

\begin{tabular}{|c|c|c|c|c|c|c|c|c|c|c|c|c|c|c|c|c|}
\hline
$X$ & $n_{sat}$  & $E_{sat}$  & $K_{sat}$  & $Q_{sat}$  & $Z_{sat}$ & $E_{sym}$ & $L_{sym}$ & $K_{sym}$ & $Q_{sym}$ & $Z_{sym}$ & $m^\star_{sat}/m$ & $\Delta m^\star_{sat}/m$ & $b$ & $\xi_\omega$ & $\xi_\rho$ & $B^\star$\\
\hline
Unit & fm$^{-3}$ & MeV & MeV  & MeV  & MeV & MeV & MeV & MeV & MeV & MeV &  & & & & & MeV\,fm$^{-3}$\\
\hline
$X_{min}$ & 0.15 & -17.5 & 190 & -1200 & -4000 & 27 & 20 & -400 & -2000 & -5000 & 0.6 & -0.1 & 1 & 0 & 0 & -20 \\
\hline
$X_{max}$ & 0.17 & -14.5 & 300 & 1000 & 5000 & 37 & 80 & 300 & 5000 & 5000 & 0.8 & 0.2 & 10 & 0.5 & 1 & 20 \\
\hline
\end{tabular}

\caption{Minimal and maximal value considered for each of the parameters of the two phases of the hybrid~EoS.}
\label{nucprior}

\end{table*}

\subsection{The NJL model}\label{sec:NJL}

We describe the deconfined quark matter phase using the three-flavor SU(3) NJL model \cite{NJL1,NJL2,Buballa,Klevansky,Hatsuda,Klimtreview}. This model has been widely used in the literature as an effective model of QCD, in hadronic physics~\cite{Klimtarticle,Rehberg_su3,Gastineau2002} and astrophysics~\cite{SchaffnerBielich1999,Menezes2003}
 alike. Its main feature is the reproduction of the symmetry properties of QCD, most notably the mechanism of spontaneous breaking of the chiral symmetry and its restoration at finite temperature and chemical potential.
The Lagrangian density of this model reads
\begin{eqnarray}\label{lagrangian}
\mathscr{L}_{NJL} &&= \overline{q}(i \slashed{\partial}-\hat{m_0})q
+ G_S\sum_{a=0}^8\Big((\overline{q}\tau^{a}q)^2 + (\overline{q}i\gamma_5\tau^{a}q)^2\Big)\nonumber\\
&& - G_\omega \Big((\overline{q}\gamma^\mu\tau^{0}q)^2+(\overline{q}\gamma^\mu\gamma^5\tau^{0}q)^2\Big) \nonumber\\
&& -G_\rho\sum_{a=1}^8\Big((\overline{q}\gamma^\mu\tau^{a}q)^2+(\overline{q}\gamma^\mu\gamma^5\tau^{a}q)^2\Big) \nonumber\\
&& -K\Big(\operatorname{det}\limits_{f}[\overline{q}(1+\gamma_5)q]+\operatorname{det}\limits_{f}[\overline{q}(1-\gamma_5)q]\Big)
\end{eqnarray}
The first term can be recognized as the relativistic free (Dirac) Lagrangian, describing the propagation of non-interacting fermions. Here, the fermions are the $u,d,s$ quarks, represented by the color triplet and flavor triplet spinor $q$. The small quark bare masses, which introduce a small explicit chiral symmetry breaking, are gathered in the diagonal matrix in flavor space ${\hat{m_0}=\diag(m_{0,u},m_{0,d},m_{0,s})}$.

The subsequent terms of Eq.(\ref{lagrangian}) describe contact interactions that emerge as the simplest way to write an interaction with only quark degrees of freedom that satisfies the flavor symmetries characterized by the group ${\text{SU(3)}_V \times  \text{SU(3)}_A \times \text{U(1)}_B}$. 
 The first three of these terms, involving the flavor Gell-Mann matrices~$\tau^a$, introduce four-quark point-like interactions in the scalar~($S$), vector-isoscalar~($\omega$) and vector-isovector~($\rho$) channels. 
All potential effects of color superconductivity, which involve four-quark coupling terms in additional channels \cite{Alford2007,Blaschke2005}, are neglected for simplicity in this exploratory work.
The last term comprising determinants in flavor space (also known as the 't~Hooft term) ensures that the axial symmetry group $U(1)_A$ is broken, mimicking the axial anomaly of QCD~\cite{KobayashiMaskawa,thooft1,thooft2,thooft3,Schafer}.

It is worth mentioning that the form of the NJL four-fermion Lagrangian can be directly derived by carrying out a Fierz transformation of a global color current-current interaction (which is itself an approximation of the QCD gauge interaction). 
Following this procedure, fundamental relationships emerge between the coupling constants in each interacting channel, yielding in particular ${G_\omega = G_\rho = G_S/2}$ \cite{Klevansky,Buballa}. 
For this reason, it is very often assumed that ${G_\omega = G_\rho = G_V}$, and the value ${G_V = G_S/2}$ is considered as a reference.
However, enforcing these relationships is not needed in order to meet the symmetry requirements of the theory.

The effective NJL interaction is meant to replace the gauge interaction coupling the quarks to the gluon dynamics in QCD. Therefore, the coupling constants we consider here ($G_S$,~$G_\omega$,~$G_\rho$,~$K$) can be interpreted to encode all the gluonic contribution of the strong interaction. 
Since the four-fermion operators have dimension 6, the NJL model is non-renormalizable. Consequently, it must be interpreted as an effective field theory, which is only valid up to a certain cutoff energy scale $\Lambda$. The parameter $\Lambda$ can also be interpreted as the scale at which the strong interaction vanishes, a crude approximation for the property of asymptotic freedom of QCD.

Note that we will use the NJL model to describe the quark matter phase, which we freely call the deconfined phase although there is no actual description of color confinement in our framework. Therefore, in the remainder of this paper, we will use interchangeably the terms \enquote{deconfinement transition} and \enquote{nuclear-quark transition} (NQ transition) to refer to the transition from nucleonic degrees of freedom to quark degrees of freedom.

In the mean field approximation, the grand potential of the NJL model can be decomposed into three terms:

\begin{equation}\label{GpotNJL}
\Omega_{NJL}=\Omega_{q}+\Omega_{V}+B_{eff}=-P_{NJL}
\end{equation}
where in the zero temperature approximation, each term reads \cite{Buballa,Pereira2016}:


\begin{eqnarray}
\Omega_{q} &&= -2N_c\sum_f\int^{\lvert \bm{p} \rvert <p_{F,f}}\frac{d^3p}{(2\pi)^3}\frac{\bm{p}^2}{3E_f} \qquad \qquad \qquad \nonumber \\
&&= -\frac{N_c}{24\pi^2}\sum_f\Big[p_{F,f}\tilde{\mu}_f(3m_f^2-2p_{F,f}^2)\nonumber \\
&&\qquad\qquad\qquad\qquad\quad
+3m_f^4\ln\Big(\frac{m_f}{p_{F,f}+\tilde{\mu}_f}\Big)\Big]
\end{eqnarray}

\begin{eqnarray}
\Omega_V &&= -\frac{2}{3}G_\omega(n_u+n_d+n_s)^2-G_\rho(n_u-n_d)^2 \qquad \nonumber\\
&&\qquad\qquad\qquad-\frac{G_\rho}{3}(n_u+n_d-2n_s)^2
\end{eqnarray}

\begin{eqnarray}
B_{eff}&&=2G_S\sum_f\langle\overline{q}_fq_f\rangle^2-4K\prod_f\langle\overline{q}_fq_f\rangle \qquad\qquad \nonumber\\
&&\qquad\qquad-2N_c\sum_f\int^{\lvert  \bm{p} \rvert<\Lambda}\frac{d^3p}{(2\pi)^3}E_f -\Omega_0
\end{eqnarray}
Here ${p_{F,f}=\Theta(\tilde{\mu}_f-m_f)\sqrt{\tilde{\mu}_f^2-m_f^2}}$ is the Fermi momentum of the quarks of flavor ${f = u, d, s}$ and ${E_f = \sqrt{\bm{p}^2+m_f^2}}$. The vacuum pressure $-\Omega_0$ is determined such that the total pressure of the model $P_{NJL}$ vanishes at zero density. The quark number densities $n_f$ and chiral condensates ${\langle\overline{q}_fq_f\rangle}$ can be calculated using:

\begin{equation}
n_f = \frac{N_c}{3\pi^2}p_{F,f}^3
\end{equation}
\begin{eqnarray}
\langle\overline{q}_fq_f\rangle&&=-\frac{N_c}{2\pi^2}m_f\Big[\Lambda \sqrt{\Lambda^2+m_f^2}-p_{F,f}\tilde{\mu}_f \qquad\nonumber \\
&&\qquad\qquad\quad+m_f^2\ln\Big(\frac{\tilde{\mu}_f+p_{F,f}}{\Lambda+\sqrt{\Lambda^2+m_f^2}}\Big)\Big]
\end{eqnarray}

The mass $m_f$ and effective chemical potential $\tilde{\mu}_f$ for each flavor are determined in the mean field approximation by a minimization of the grand potential with respect to these parameters, which yields

\begin{equation} \label{CSB}
m_i = m_{i,0}-4G_S\langle\overline{q}_iq_i\rangle +2K\langle\overline{q}_jq_j\rangle\langle\overline{q}_kq_k\rangle
\end{equation}
\begin{equation}
\tilde{\mu}_i=\mu_i-\frac{4}{3}\Big(G_\omega(n_i+n_j+n_k)+G_\rho(2n_i-n_j-n_k)\Big)
\end{equation}
where $i,j,k$ denotes any permutation of the $u,d,s$ flavors. The former equation illustrates the mechanism of spontaneous chiral symmetry breaking in the NJL model, through which quarks acquire a dynamical mass proportional to the chiral condensates (plus a small contribution due to the bare quark masses). The latter expresses the effects of vector interactions: quarks obtain an effective chemical potential $\tilde{\mu}_f$ which is shifted to a lower value than the (physical) chemical potential~$\mu_f$. Once all these quantities have been fixed, the energy density of the model can be easily computed with the usual thermodynamic formula:
\begin{equation}
\rho_{NJL}=-P_{NJL}+\sum_{f}\mu_f n_f
\end{equation}

In the NJL model, vacuum parameters are fixed to reproduce known properties of low-energy hadronic physics such as the meson mass spectrum. The resulting values of this fit, taken from~\cite{Pereira2016}, are given in Table~\ref{NJLparam}. 
In~this work however, the repulsive vector couplings in the isoscalar ($\omega$) and isovector ($\rho$) channels are introduced as free parameters, as they have been reported to be crucial in order to stiffen the EoS enough for hybrid stars to reach the $2M_\odot$ threshold \cite{Pereira2016,Sedrakian,Masuda2013}.
This can be further justified as there are QCD effects, neglected in the NJL effective model approximation, that would lead to an in-medium modification of the coupling constants (for example, instanton effects at high temperature can affect the effective vector coupling constant \cite{Schafer:1994nv}). Hence it is usually accepted to treat those parameters as free when studying dense quark matter.

We use as parameters the vector to scalar coupling ratios ${\xi_\omega=\frac{G_\omega}{G_S}}$ and ${\xi_\rho=\frac{G_\rho}{G_S}}$, which we vary considering a flat distribution in the intervals $[0.0,0.5]$ and $[0.0,1.0]$, respectively. 
In \cite{Pereira2016,Ferreira2020,Pagliara_2008,Sedrakian}, it was also suggested that an additional parameter $B^\star$ could be introduced to increase the freedom one has on the quark EoS parametrization. 
This parameter, very analogous to the bag constant in the MIT bag model of QCD, simply shifts the pressure and energy density of the quark phase without affecting other thermodynamic quantities, following:
\begin{equation}
P_Q = P_{NJL}+B^\star
\end{equation}
\begin{equation}
\rho_Q = \rho_{NJL}-B^\star.
\end{equation}

In general, the bag constant is a pressure characteristic of the QCD vacuum, related to its non-trivial structure, for instance due to the presence of a quark condensate in the hadronic phase. 
In principle, $B^\star$ can be computed within a given model (here by fixing the value of $\Omega_0$ above), but the NJL model is missing the QCD gluonic degrees of freedom which should also contribute to the bag pressure. It is therefore justified to allow deviations from the NJL prescription and take this parameter as free.
The effective bag constant $B^\star$ is added to the parameter space of the quark phase, with a flat distribution in the interval $[-20$ MeV\,fm$^{-3},\,B_{max}]$, where the value of $B_{max}$ depends on the hadronic and quark models, and is discussed in Sec. \ref{sec:Bstar} below. 
We denote by $\mathbf{X}$ the entire hybrid parameter space, given by the reunion of the nuclear parameters $\mathbf{X}_N$ and the three free parameters in the quark sector $\{\xi_\omega,\xi_\rho,B^\star\}$.

\begin{table}[htbp]
\centering
\begin{tabular}{|c|c|c|c|c|c|c|}
\hline
 $m_{0,u}$ (MeV)  & $m_{0,d}$ (MeV) & $m_{0,s}$ (MeV)  & $\Lambda$ (MeV) & $G_S\Lambda^2$ & $K\Lambda^5$ \\
\hline
 5.5 & 5.5 & 135.7 & 630 & 1.781 & 9.29\\
\hline
\end{tabular}
\caption{Model parameters used for the quark phase taken from~\cite{Pereira2016}. These parameters are assumed fixed in our EoS prior.}
\label{NJLparam}
\end{table}

\subsection{The nuclear-quark phase transition}

\subsubsection{Maxwell versus Gibbs construction}

We assume that the deconfinement phase transition transition is first order and connect the two phases via the Maxwell construction. The phase transition point is calculated by assuming mechanical and baryo-chemical equilibrium between the two phases:

\begin{equation} \label{equilibrium}
P_N = P_Q \,,\quad \quad \quad \quad \quad \mu_N = \mu_Q
\end{equation}
Since we work under the zero temperature approximation in both phases, thermal equilibrium is also automatically fulfilled. 
It was pointed out by Glendenning~\cite{Glende} that in the one-dimensional Maxwell construction, where the order parameter is given by the baryonic charge $n_B$ and the charge neutrality condition $n_C=0$ is enforced separately in the two phases, charge equilibrium cannot be achieved because the electron chemical potentials are in principle different in the two phases: ${\mu_{e,N}\neq\mu_{e,Q}}$.
According to this argument, thermodynamic consistency would rather require the separate equality of the two chemical potentials $\mu_B$ and $\mu_C$  (or equivalently, the neutron and electron chemical potentials $\mu_N$ and $\mu_e$) within a two-dimensional Gibbs construction involving the two independent conserved charges $n_B$ and $n_C$~\cite{Glende,favored,global}. 
However, it was argued by Chomaz et al.~\cite{Chomaz1,Chomaz2} that the long range character of the electromagnetic interaction implies that the total charge cannot be associated with a chemical potential, even if it is a conserved quantity. This reduces the dimensionality of the order parameter and guarantees the thermodynamic consistency of the Maxwell construction.

In the core of a neutron star, the transition between hadronic and deconfined matter might well occur through an inhomogeneous phase consisting of coexisting individually charged domains of (less dense) hadronic and (more dense) quark matter \cite{Glendenning1995,Glendenning2001,Xia2020,Schmitt2020,Ju2021,Maslovpasta}. 
However, because of the divergence of the Coulomb energy density at the thermodynamic limit if the net electric charge is not zero, those domains must be mesoscopic and the interface energy not negligible, thus preventing a phase coexistence ruled by a standard Gibbs construction. 
It was suggested that the energy balance of the interface should favor the presence of the mixed inhomogeneous phase, thus quenching the first order phase transition~\cite{Maruyama}. However, the finite size effects at the interface between quark and hadronic matter are directly proportional to the surface tension between the two phases. Since the value of this surface tension is unknown to this day (and in addition should in principle depend on the model used to describe both phases), the exact structure of the mixed phase is very much uncertain. Interestingly, in the limit of very high surface tension, the structured mixed phase becomes energetically disfavored and the Maxwell picture is recovered.

For these reasons, we stick to the simple first order phase transition ruled by the Maxwell construction in the present study.

\subsubsection{The role of $B^\star$}\label{sec:Bstar}

As introduced in Sec. \ref{sec:NJL}, besides the freedom one has with the vector coupling constant, the quark equation of state can gain some extra flexibility  by introducing the effective bag constant parameter 
$B^\star$. The determination of a physically relevant interval for the value of this parameter requires some discussion, that we now address. 
 
\begin{figure}[htbp]
\includegraphics[scale=0.68]{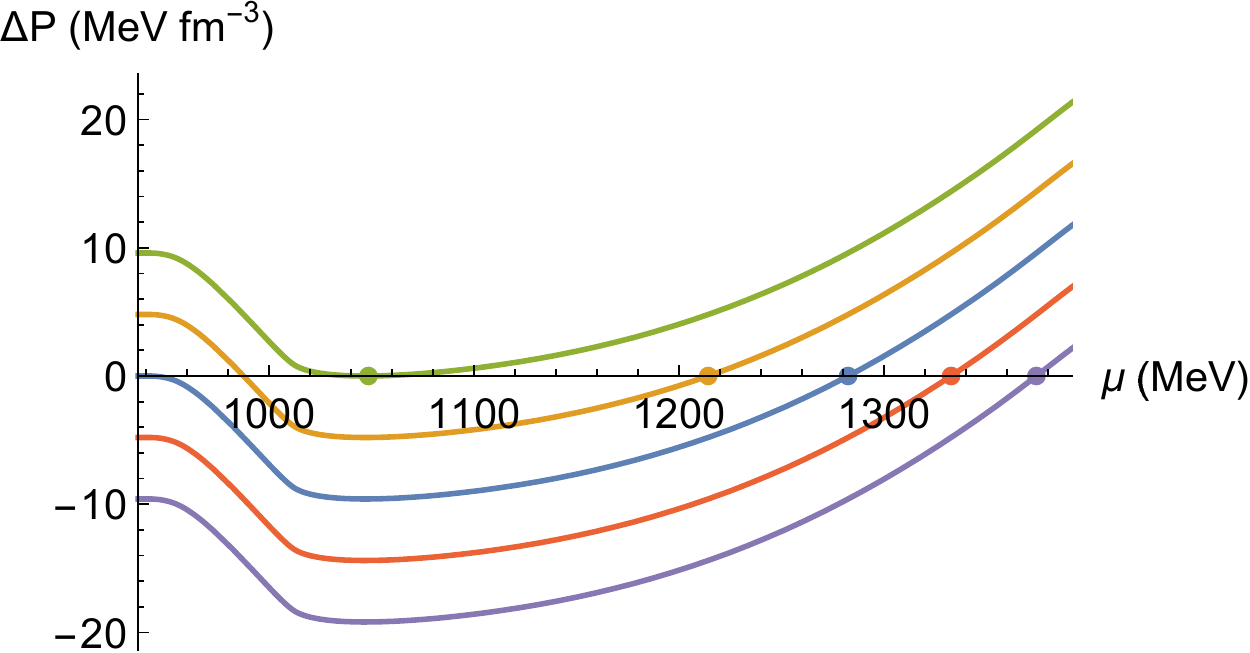}
\caption{Pressure difference ${\Delta P=P_Q-P_N}$ as~a function of baryon chemical potential $\mu$ for different values of the  parameter $B^\star$ (from bottom to top: $B^\star$=$-B_{max},\,-B_{max}/2,\,0,\, B_{max}/2, \,B_{max}$). The nuclear model is the meta version of DDME2 \cite{DDME2}, while the quark model uses $\xi_\omega=0.25$ and $\xi_\rho=0$, for which ${B_{max}=9.59}$~MeV\,fm$^{-3}$. The dots indicate the chemical potential of the phase transition for each value of $B^\star$.}
\label{pressdif}
\end{figure}

The effect of $B^\star$ is very clear: if positive, it reduces the free energy of the quark phase and therefore pulls back the transition density, pressure, and chemical potential to lower values. If negative, the opposite behavior should be observed. 
This effect can be seen on Fig.\ref{pressdif}, where the pressure difference between the quark phase and nuclear phase ${\Delta P = P_Q - P_N}$ is plotted as function of the chemical potential for different values of $B^\star$. The other parameters are chosen arbitrarily for illustrative purposes. We recall that in the grand canonical ensemble, the thermodynamically preferred phase has the lowest free energy $\Omega=-P$, or equivalently the highest pressure. Therefore, if ${\Delta P<0~(>0)}$ the nuclear (quark) phase is the most stable. The phase transition occurs when ${\Delta P=0}$.

In Fig.\ref{pressdif}, we indeed observe the expected behaviors depending on the sign of $B^\star$. If $B^\star$ is negative, $\Delta P$ is negative at small chemical potentials and eventually becomes positive if $\mu$ reaches high enough values. In principle, $B^\star$ can be lowered arbitrarily towards the negative values; this will simply push the phase transition to higher and higher densities, such that eventually quark matter never appears in stable NS configurations. Conversely, if $B^\star$ is positive, the phase transition occurs earlier as expected with the NJL prescription. In addition to this obvious effect, another feature is visible in Fig.\ref{pressdif}:  we can see that for very low chemical potentials the quark phase becomes thermodynamically favored over the nuclear phase when $B^\star>0$. This is of course unphysical, and only a transition for which $\Delta P$ goes from a negative value (nuclear phase favored) to a positive value (quark phase favored) can be considered as physically meaningful. 
Therefore, if $B^\star>0$ we assume that the nuclear phase is favored until $\Delta P$ reaches positive values a second time. This also means that $B^\star$ cannot exceed a limiting value, denoted here as $B_{max}$ and defined by:
\begin{equation}
B_{max}=\Big\lvert\min_{\mu}(P_{NJL}-P_N)\Big\rvert
\end{equation} 
If ${B^\star>B_{max}}$, the quark phase would be thermodynamically favored over all possible densities, which is of course inconsistent with our knowledge of low-energy nuclear physics. 
If ${B^\star=B_{max}}$, we also notice that the density discontinuity of the phase transition~$\Delta n$ (which corresponds to the slope $\frac{\partial \Delta P}{\partial \mu}$ at the phase transition point on Fig.\ref{pressdif}) vanishes.
The value of~$B_{max}$ depends very much on the nuclear and quark parameters chosen, but typically lies in the range {5-100 MeV\,fm$^{-3}$}.

In~\cite{Pagliara_2008,Pereira2016} it was suggested that the value of $B^\star$ could be tuned so as to enforce the deconfinement transition to occur simultaneously with chiral symmetry restoration in the NJL model. The chemical potential of the chiral phase transition $\mu_\chi$ was determined explicitly if the transition was first order, or by a maximization of the chiral condensate susceptibilities if the transition was a crossover (which typically happens with finite values of the vector couplings). Then the corresponding value of $B^\star$ was calculated such that $\Delta P=0$ at $\mu=\mu_\chi$:
\begin{equation}
B^\star_\chi=P_N|_{\mu=\mu_\chi}-P_{NJL}|_{\mu=\mu_\chi}
\end{equation}
Because the chiral phase transition usually happens at relatively low density \footnote{The chirally broken phase is composed of quarks with high effective mass, which therefore do not contribute much to the density or pressure. Only beyond the chiral symmetry restoration does the density start to increase significantly. Typical transition densities lie in the range $1-2$ times nuclear saturation density for our parametrization.}, $B^\star_\chi$ has to take a positive value to reduce the density of the deconfinement transition. By construction, we always have $B^\star_\chi\leq B_{max}$ since we do impose $\Delta P = 0$ at some chemical potential, which could not be achieved otherwise. 
However, this construction is not always consistent, since we  cannot guarantee that $\mu_\chi$ does not correspond to the unphysical $\Delta P=0$ solution pictured in Fig.\ref{pressdif} (see the yellow line second from the top with $B^\star = B_{max}/2$).
In other words,  we do not impose that the baryon density of the quark phase is higher than that of the nuclear phase at $\mu_\chi$.
Therefore, even though choosing $B^\star=B^\star_\chi$ is possible, it does not always end up making the chiral transition and the deconfinement transition simultaneous. 
For a specific set of models, this method might be suitable, because it may work in some cases, but it cannot be applied systematically to a large number of randomly generated models for a Bayesian analysis. 
We observed that this method typically fails if vector interactions are high enough, because they tend to increase the range of chemical potentials in which ${\frac{\partial \Delta P}{\partial\mu}<0}$ and make the chiral transition smoother.

In~\cite{Ferreira2020,Sedrakian}, $B^\star$ was taken as a free parameter, taking positive values of the order of $10$~MeV\,fm$^{-3}$. The restriction ${B^\star\leq B_{max}}$ was not mentioned. In the following, we will adopt two different hypothesis. In the conservative hypothesis, which we call $H_A$, we stick to the original NJL prescription and keep $B^\star=0$. In the second one, called $H_B$, we add $B^\star$ to our set of model parameters for the prior equations of state and choose a prior distribution range of $[-20,20]$ MeV\,fm$^{-3}$, while always enforcing the physical limitation  ${B^\star < B_{max}}$. The range is centered around zero, as there is a priori no clear theoretical reason to prefer positive or negative values.

\subsection{Bayesian analysis}
In this section, we describe the different steps that were followed in order to obtain the posterior distributions of our Bayesian analysis. The results will be presented in the next section.

First, we generate a prior sample for the parameters of both the nuclear and the quark model. On the nuclear side, a set of $\approx 10^8$ models is produced with empirical parameters generated randomly following flat and uncorrelated distributions in the ranges defined in Table \ref{nucprior}.
On the quark side, 121 different EoSs are generated under $\beta$-equilibrium and charge neutrality conditions, with 11 different evenly spaced values for each vector parameters in the intervals ${\xi_\omega\in[0.0,0.5]}$ and ${\xi_\rho\in[0.0,1.0]}$. 
We choose to use a grid for the distributions of $\xi_\omega$ and  $\xi_\rho$ instead of a random flat prior for practical reasons: since the quark EoS are already computed beforehand, we gain significant computation time not having to generate a new one for each hybrid model considered. 
As the quark EoS has fewer parameters than the nuclear one in our framework, we do not require as many different models to explore the whole parameter space. 
The very large nuclear sample is motivated by the fact that stringent theoretical constraints exist on the behavior of the energy per particle of pure neutron matter from many-body perturbation theory using chiral effective interactions~\cite{Drischler,chiral}.
We will include these constraints as a pass-band filter following previous works~\cite{Carreau_2019,Hoa1,Hoa2} which will sensibly reduce the size of our sample to about 10000 models, the minimum statistics needed for a convergence of the results presented in this paper.
On the other side, we have checked that the limited number of quark EoS is sufficient to cover all the possible behaviors of the quark branch within the reduced set of parameters of the NJL model.
The difference between the statistics in the two phases can be further understood from the fact that we want to explore the parameter space associated with the existence of a quark core, and it is expected that this depends on the stiffness of the hadronic branch of the EoS. 
Since the latter is modeled by a polynomial expansion in density with both positive and negative contributions, a large sample is needed to ensure the convergence of the predictions. 

To ensure the compatibility of the parameter set with the \textit{ab initio} calculation of low density nuclear matter,  
for ten evenly spaced densities in the range $[0.02,0.2]$~fm$^{-3}$, the energies per particle of the model for both symmetric and pure neutron matter are compared to the corresponding chiral effective field theory ($\chi$EFT) energy bands from~\cite{Drischler}. The nuclear model is then rejected if it predicts an energy more than 5\% away of the uncertainty of the \textit{ab initio} calculation.  

For each successful set of nuclear parameters, the EoS of neutron star matter (\textit{i.e.},~with $\beta$ equilibrium and charge neutrality) is then calculated, including the inhomogeneous phase in the outer and inner crust. For the crust calculation, a fit of the surface parameters $\textbf{X}_\sigma$ is performed on the nuclear mass table of AME2012, thus guaranteeing a consistency between bulk and surface properties. The computation is carried out with increasing density until no solution to the $\beta$-equilibrium condition is found, or if the model breaks one of the following thermodynamic consistency conditions:
\begin{enumerate}[label=(\roman*)]
\item $0\leq c_s^2 \leq 1$
\item $\frac{dP}{dn_B}>0$
\item $e_{sym}=\frac{1}{n_B}\frac{\partial\rho}{\partial\delta^2}\Big|_{n_B}>0$
\end{enumerate}
In the quark phase, the previous conditions are always met by construction. In this way, we ensure that we always consider hybrid EoS that are thermodynamically consistent. If one of the previous conditions is broken before the EoS reaches $2n_{sat}$, the model is discarded since we expect nuclear matter to be stable until at least~${\approx 2n_{sat}}$. Note that, anticipating our results, a transition to quark matter below $2n_{sat}$ is highly unlikely anyway in our model.

To build the hybrid EoS, we then attempt to carry out a Maxwell construction with each of the 121 quark EoS. In the case of hypothesis $H_B$, a bag parameter $B^\star$ is added to the quark EoS, randomly generated in the range $[-20,20]$ MeV\,fm$^{-3}$ for each pair of models.
If no phase transition can be constructed  (either because the phase transition is located in a region where the nuclear EoS is not thermodynamically consistent anymore, or because $B^\star>B_{max}$), the hybrid model candidate is discarded. 

For each hybrid EoS obtained, we solve the Tolman-Oppenheimer-Volkoff (TOV) equations for spherical non rotating stars \cite{Tolman,OV} and compute the maximum mass $M_{TOV}$ associated with each EoS. In order to account for the observation of pulsar masses above 2$M_\odot$ (where $M_{\odot}$ is the solar mass), models which cannot sustain stars with 2$M_\odot$ are discarded. 
We have checked that implementing a likelihood filter based on the mass measurement from radio-timing observations of pulsar PSR J0348+0432 \cite{Antoniadis2013}, ${M_{J0348} = 2.01 \pm 0.04 M_{\odot}}$, instead of the rough elimination of models with ${M_{TOV}<2M_\odot}$, does not change the results presented below.
Finally, if the maximum pressure achievable by the model $P_{TOV}$ is lower than the pressure of the phase transition $P_t$, we also discard the model as it would mean quark matter cannot be reached in compact stars.
In this way, we only consider hybrid models for which hybrid stars are possibly realized.
The whole set of successful hybrid EoSs constitutes our prior sample.

The posterior distributions of observables $\mathcal{O}(\bf X)$ that can be computed given the set $\bf X$ of parameters of the hybrid EoS  are conditioned by likelihood models of the different observations and constraints $c_k$ according to the standard definition
 
\begin{equation}
	P(\mathcal{O}|{\mathbf c})=\mathcal N   P(\mathcal{O}) \prod_k w_k(c_k)\,,
\end{equation}
where $\mathcal N$ is a normalization factor, and $ P(\mathcal{O})$ is the prior distribution:

\begin{equation}
 P(\mathcal{O})=\frac{1}{N_{tot}}\sum_{i=1}^{N_{tot}}\delta(\mathcal{O}(\bf X_i)-\mathcal{ O})  
\end{equation}
with $N_{tot}$ the total number of prior models.
 
Two different constraints $c_k$ are used in the present study, coming from low energy nuclear physics and  astrophysical observations respectively.

The first constraint concerns the quality of reproduction of the $N_N=2149$ nuclear mass measurements compiled in the AME2012 mass evaluation~\cite{AME2012}:
 
\begin{equation}
w_{AME}=\frac{1}{N}\exp\Big(-\frac{\chi_{AME}^2}{2}\Big) ,
\end{equation}
where $\chi_{AME}$ is the evaluation of the quality of the fit,
\begin{equation}
\chi_{AME}^2=\frac{1}{\nu}\sum_{i=1}^{N_N}\bigg(\frac{M^{(i)}-M_{AME}^{(i)}}{\sigma^{(i)}}\bigg)^2 .
\end{equation}

Here, $M_{AME}^{(i)}$ and  $M^{(i)}$  represent respectively the experimental  and theoretical nuclear masses, the latter being calculated within a compressible liquid drop model approximation (Eq.\ref{eq:enuc}) using the best-fit surface and curvature parameters for each EoS model (see Sec.~\ref{sec:meta} above); 
 $\sigma^{(i)}$ represents the systematic theoretical error, and  ${\nu=  N_N-4}$ is the number of degrees of freedom. 

 Second, we add another layer of filters by confronting the tidal deformability predictions of each model to the results obtained by the LIGO-VIRGO Collaboration (LVC) using the gravitational wave measurements from the GW170817 event~\cite{Abbott19}. From the analysis of the signal, the probability distribution functions (PDFs) of the binary tidal deformability $\tilde{\Lambda}$ and of the lowest mass component $M_2$ were extracted. 
 The associated likelihood weight $w_{GW170817}$ reads

\begin{equation} \label{GWweight}
w_{GW170817} = 
\sum_{j}p_{LVC}^{(M)}\Big(M_2^{(j)}\Big)p_{LVC}^{(\Lambda)}\Big(\tilde{\Lambda}^{(j)}\Big)  
\end{equation}
Here, the $p_{LVC}$ functions are the PDFs taken from the LVC data, and $\Big(M_2^{(j)}\Big)_j$ is a set of mass values that was chosen to probe the entire likelihood range of $M_2$. In this work, we used an evenly spaced grid of 27 points between 1.1 and 1.36 $M_\odot$. Since we know very precisely the chirp mass of the system ($\mathcal{M}=1.186\pm0.001M_\odot$), we can determine the mass $M_1$ of the heaviest object by inverting the following relationship, neglecting the chirp mass uncertainty:
\begin{equation}
\mathcal{M}=\frac{(M_1M_2)^{\frac{3}{5}}}{(M_1+M_2)^{\frac{1}{5}}}\,,
\end{equation}
which yields the result
\begin{equation}
M_1 = y(\mathcal{M},M_2)+\frac{\mathcal{M}^5}{3M_2^3 y(\mathcal{M},M_2)}\,,
\end{equation}
\begin{equation}
y(\mathcal{M},M_2) = \bigg(\frac{\frac{\mathcal{M}^5}{M_2^2}+\sqrt{\frac{\mathcal{M}^{10}}{M_2^4}-\frac{4\mathcal{M}^{15}}{27M_2^9}}}{2}\bigg)^{\frac{1}{3}}\,.
\end{equation}

Then, for the values of $M_1^{(j)}$ and $M_2^{(j)}$, we calculate the associated tidal deformabilities $\Lambda_1^{(j)}$ and $\Lambda_2^{(j)}$ for each model solving the usual equations from general relativity \cite{TidalGR,LoveGR}. The binary tidal deformabilities are then calculated using the formula

\begin{equation}
\tilde{\Lambda}=\frac{16}{13}\frac{(M_1+12M_2)M_1^4\Lambda_1+(M_2+12M_1)M_2^4\Lambda_2}{(M_1+M_2)^5}\,,
\end{equation}
which eventually allows us to compute the weight of Eq.(\ref{GWweight}) for each model considered.

We can observe that Eq.(\ref{GWweight}) supposes statistical independence between $\tilde{\Lambda}$ and the mass $M_2$ of the lower component.
It was checked in~\cite{Hoa2} that the results are not modified if we consider instead the full joint posterior distribution of $\tilde{\Lambda}$ and $q = M_2/M_1 $ from~Ref.\cite{Abbott19}.

With an initial sampling of $10^8$ nuclear models, the posterior size is of around 100 000 models, with around 3000 different nuclear models surviving all the filters and producing at least one viable hybrid EoS. This makes an average of about 30 hybrid models per nuclear model.

\section{Results} \label{sec::results}

In this section are gathered all the results obtained from the analysis of the posterior distributions. A~summary of the numerical results for each quantity examined is shown in Table \ref{Meanvalues} in the Appendix.

\subsection{Model parameters posteriors}
\subsubsection{Quark sector}

\begin{figure}[htbp]
\includegraphics[scale=0.28]{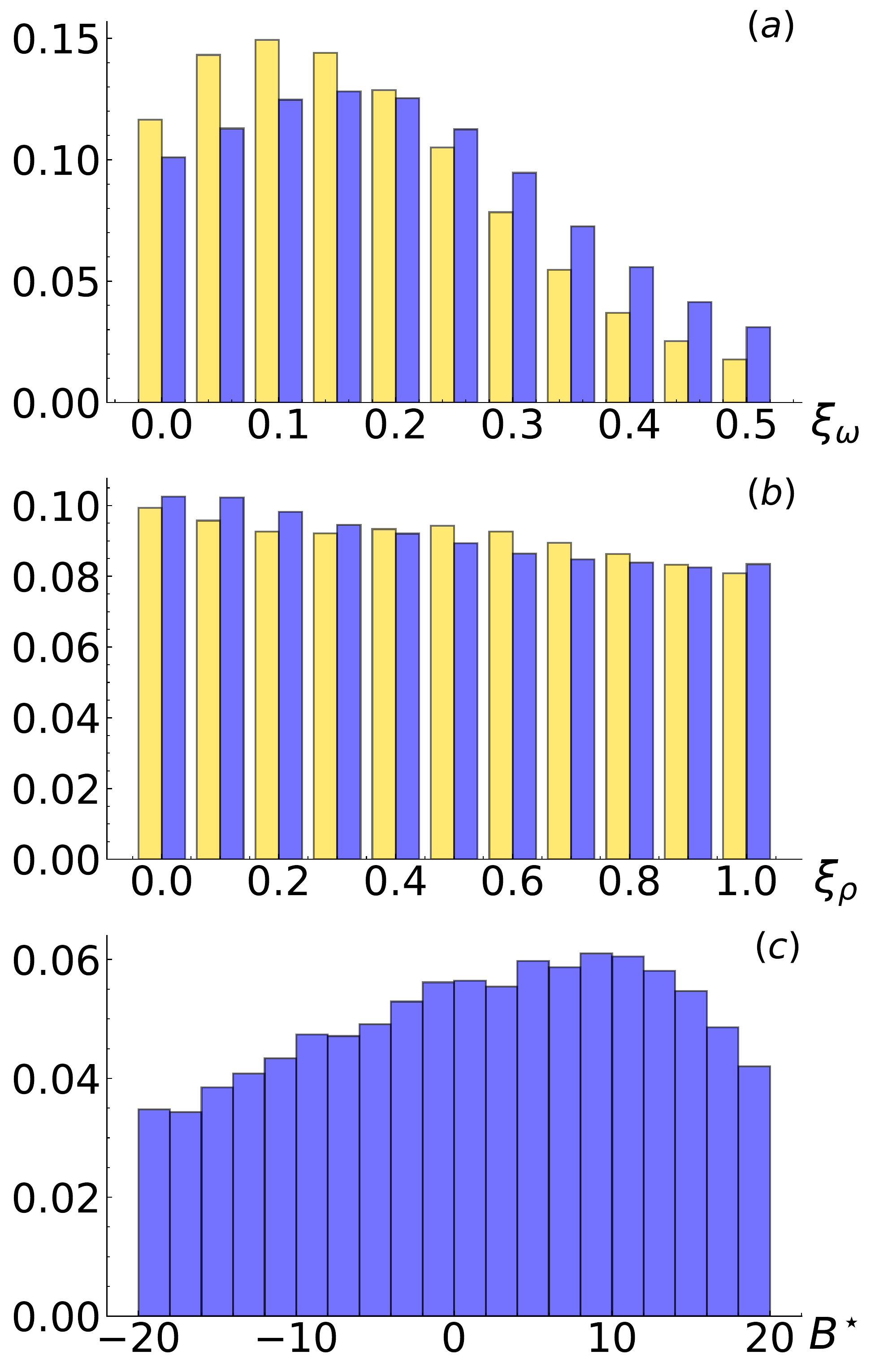}
\caption{Top and middle: Posterior distributions of the NJL vector interaction couplings $\xi_\omega$ and $\xi_\rho$ 
with hypotheses $H_A$ (yellow on the left) and $H_B$ (blue on the right) (see text for details). Bottom: Posterior distribution of the bag parameter $B^\star$ (in~units MeV\,fm$^{-3}$) with hypothesis $H_B$. Each histogram is normalized such that the sum of each bin height is equal to~1.}
\label{quarkparams}
\end{figure}

First, we show on Fig.\ref{quarkparams} the posterior distributions of the parameters of the quark EoS : the vector couplings $\xi_\omega$, $\xi_\rho$ and the effective bag constant $B^\star$. In the case of the vector couplings, two different distributions are obtained depending on the hypothesis made on the parameter $B^\star$, namely $H_A$ for which $B^\star=0$ (yellow) and $H_B$ for which $B^\star$ is allowed to vary (blue); see Sec.~\ref{sec:Bstar} for details.
Interestingly, we can see that the presence or not of this extra parameter only marginally influences the values of the couplings that lead to hybrid stars.
In both cases, we observe that the distribution of the $\omega$ coupling has a relatively wide peak around ${\xi_\omega\approx 0.15}$. 
On one hand, low values of $\xi_\omega$ result in a relatively soft quark EoS at high densities which will struggle to reach high $M_{TOV}$ (unless the threshold of 2$M_\odot$ has already been reached before the PT). 
In addition, the $\omega$ channel tends to push the phase transition to larger densities, such that low values of $\xi_\omega$ are typically associated with stars with sizable quark cores, but unable to reach high mass and are therefore discarded in our posterior.
On the other hand, high values of $\xi_\omega$ stiffen the quark EoS too much, which pushes the phase transition to too high densities. The latter effect is, however, slightly reduced by positive values of the bag pressure $B^\star$, which decrease the density of the PT while not affecting the stiffness of the EoS. 
In the case of the $\rho$ channel, the posterior distributions of the coupling are as flat as the prior, indicating that (i) the possibility of an hybrid star is essentially ruled by the vector-isoscalar coupling $\omega$, and (ii) the mass and tidal polarizability measurements are not sensitive to this coupling, which is mainly linked to the flavor composition of the deconfined matter.
While essentially only positive values of $B^\star$ were previously considered in the literature \cite{Pagliara_2008,Pereira2016,Ferreira2020,Sedrakian}, we can see from the lower panel of Fig.\ref{quarkparams} that negative values of $B^\star$ do not strongly hinder the possibility of hybrid stars:
with a flat prior in the range $[-20,20]$ MeV\,fm$^{-3}$, $B^\star$~remains relatively well distributed, although an overall slight preference for positive values is observed.  Values that are too large tend to be rejected more often as they are more likely to exceed $B_{max}$, while too low values might push the PT to unreasonable densities.

\begin{figure}[htbp]
\includegraphics[scale=0.42]{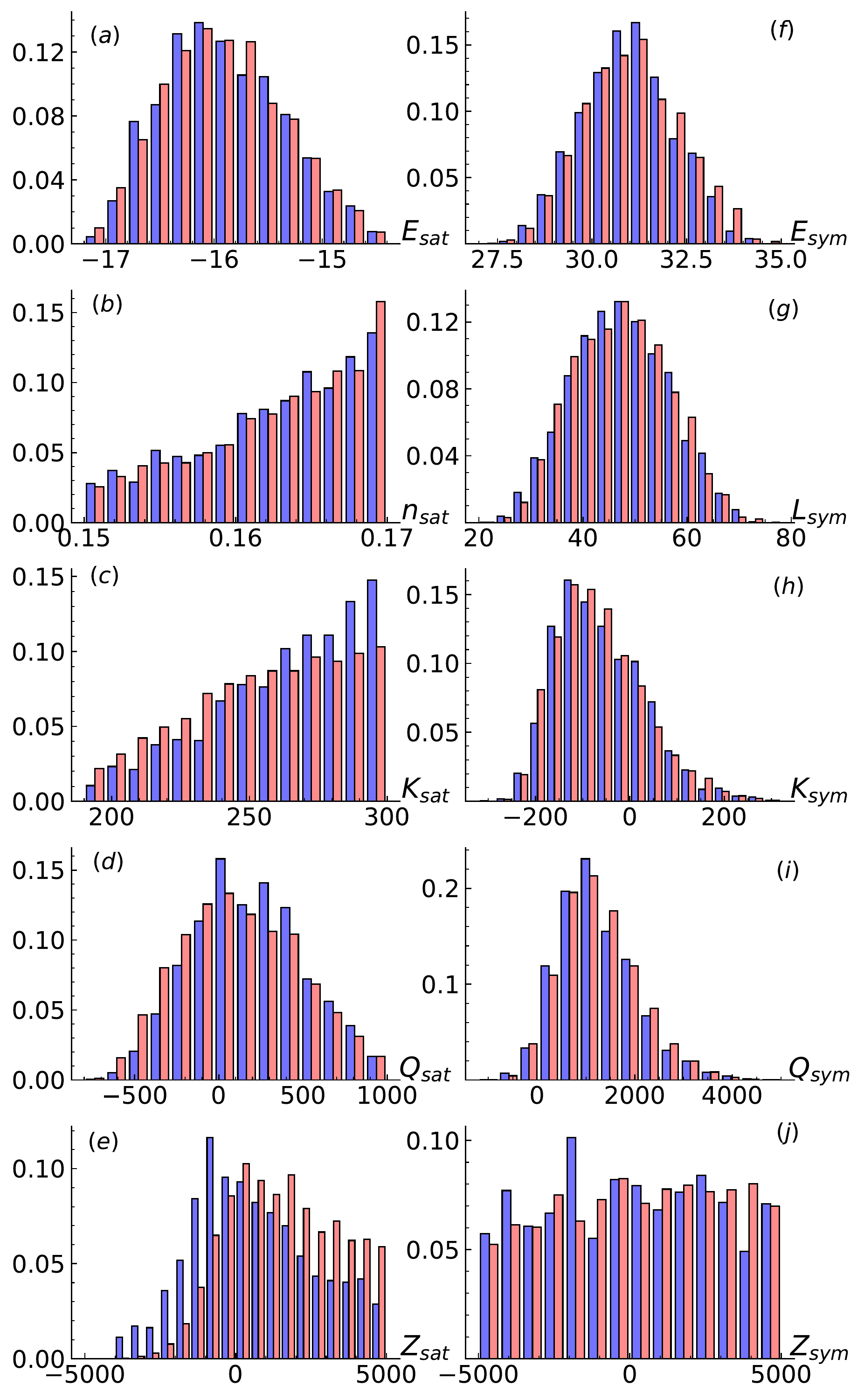}
\caption{Posterior distributions of the isoscalar (left column) and isovector (right column) nuclear empirical parameters. On each panel, the blue histogram on the left corresponds to the posterior distribution for hybrid stars under the hypothesis $H_B$, while the red one on the right is associated with a purely nucleonic EoS. Same normalization as Fig. \ref{quarkparams}.}
\label{NEP}
\end{figure}

\begin{table*}[htbp]
\centering

\begin{tabular}{|c|c|c|c|c|c|c|c|c|c|c|}
\hline
$X$ & $n_{sat}$  & $E_{sat}$  & $K_{sat}$  & $Q_{sat}$  & $Z_{sat}$ & $E_{sym}$ & $L_{sym}$ & $K_{sym}$ & $Q_{sym}$ & $Z_{sym}$ \\
\hline
Unit & fm$^{-3}$ & MeV & MeV  & MeV  & MeV & MeV & MeV & MeV & MeV & MeV  \\
\hline
Average hybrid & 0.163 & -15.9 & 263 & 177 & 656 & 30.9 & 47.4 & -62.8 & 1211 & 11  \\
\hline
Average nuclear & 0.163 & -15.9 & 255 & 105 & 1620 & 31.0 & 46.9 & -70.5 & 1225 & 297  \\
\hline
$K_{KS}$ & 0.029 & 0.024 & 0.137 & 0.110 & 0.217 & 0.042 & 0.027 &  0.047 & 0.032 & 0.058 \\
\hline
\end{tabular}

\caption{Results of the Kolmogorov-Smirnov statistical test used to compare the posterior distributions with two different hypotheses on the content of the star (hybrid-$H_B$ and purely nuclear).}
\label{KStest}

\end{table*}

\subsubsection{NEP}

In Fig.\ref{NEP}, we show histograms of the posterior distributions of the ten nuclear empirical parameters (NEPs) controlling the behavior of the hadronic EoS. For each parameter, the distribution obtained in the hybrid star hypothesis (with $H_B$) is compared to the one corresponding to the hypothesis that no phase transition occurs, \textit{i.e.} with stars entirely made of nucleonic matter.
Therefore, the difference between the two distributions quantifies how much the hypothesis of the existence of a nuclear-quark PT at high density can affect our conclusions on these parameters. Somehow not surprisingly, we see that the lower order NEPs ($n_{sat},E_{sat},E_{sym},L_{sym}$) are only weakly affected by the hypothesis, if at all. Indeed, these parameters only drive the behavior of the EoS around saturation density and do not influence the high density behavior where quarks may be involved. 
More surprisingly, the high order parameters in the asymmetry sector are also negligibly influenced by the condition that a PT occurs. 
Only the isoscalar sector is slightly affected: the average values of both $K_{sat}$ and $Q_{sat}$ are indeed increased by the assumption of a deconfinement PT, while that of $Z_{sat}$ is decreased. This effect is easy to understand. For a given quark EoS, quarks tend to appear at lower densities (and are hence favored) if the nuclear EoS is stiffer. Therefore, stiff nuclear EoSs (\textit{i.e.}, with high values of $K_{sat}$ and $Q_{sat}$) are more likely to be compatible with a quark EoS to form a viable hybrid EoS. However, in order to compensate this effect, the values of $Z_{sat}$ must be lowered in order to keep a causal and thermodynamically stable EoS, and further satisfy the constraints from $\chi$EFT. 

For a more quantitative comparison of the distribution, we gathered in Table~\ref{KStest} the mean values obtained for each NEP and for both assumptions. We also calculated the Kolmogorov-Smirnov statistic $K_{KS}$ associated with each pair of statistical distributions displayed in Fig.\ref{NEP}, defined for the quantity $X$ by:
\begin{equation}
K_{KS}(X) = \sup_{x} \Big\lvert F_{Q,X}(x) - F_{N,X}(x) \Big\rvert
\end{equation}
where $F_{Q,X}$ (resp. $F_{N,X}$) is the empirical distribution function associated with the posterior distribution of parameter $X$ in the hybrid (resp. nuclear) hypothesis. A large value of $K_{KS}$ indicates that the distributions are very different, while a low value indicates the distributions are similar. This analysis confirms the visual findings of Fig.\ref{NEP}: only $K_{sat}$ and $Z_{sat}$ are significantly influenced by the requirement that a PT takes place.

It is interesting to observe that present laboratory constraints from the energy of the giant monopole resonance excitation point towards a compressibility value $K_{sat}=230\pm 20$ MeV \cite{Khan2012,Margueron_2018}. This would suggest that the existence of hybrid stars is 
disfavored by the present empirical knowledge, at least in the hypothesis that the quark phase is satisfactorily described by the NJL model.

\begin{figure}[htbp]
\includegraphics[scale=0.25]{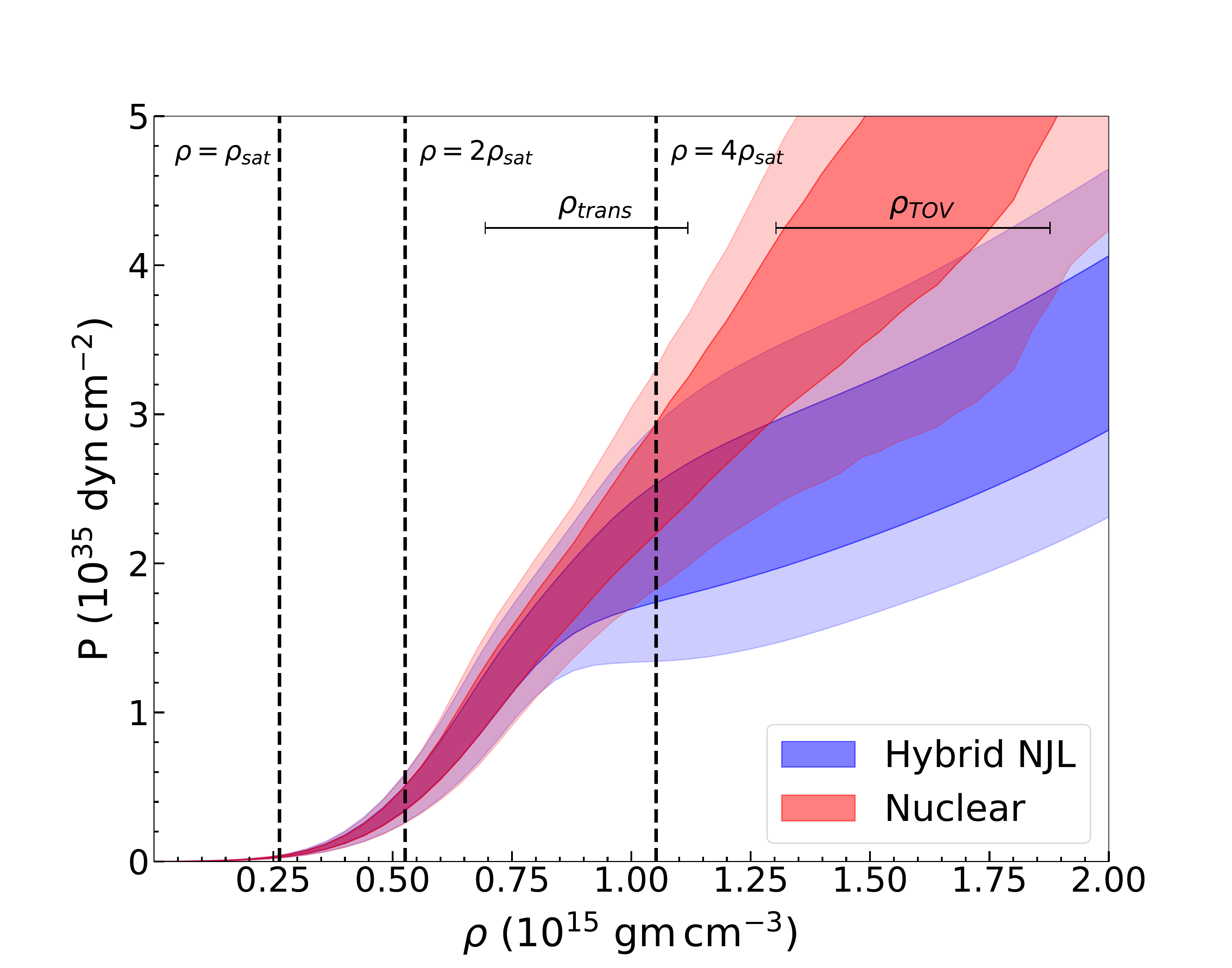}
\caption{Comparison of the posterior EoS in the hypothesis of presence (blue) or absence (red) of PT. The darker (resp lighter) regions correspond to 1$\sigma$ (resp 2$\sigma$) uncertainty areas.
The average and 1$\sigma$ uncertainty on the NQ transition density $\rho_{t}$ as well as on the central density of the maximum mass configuration $\rho_{TOV}$ (in the hybrid hypothesis) are also displayed for reference.}
\label{compEOS}
\end{figure}

\subsection{Equation of state}

\begin{figure}[htbp]
\includegraphics[scale=0.25]{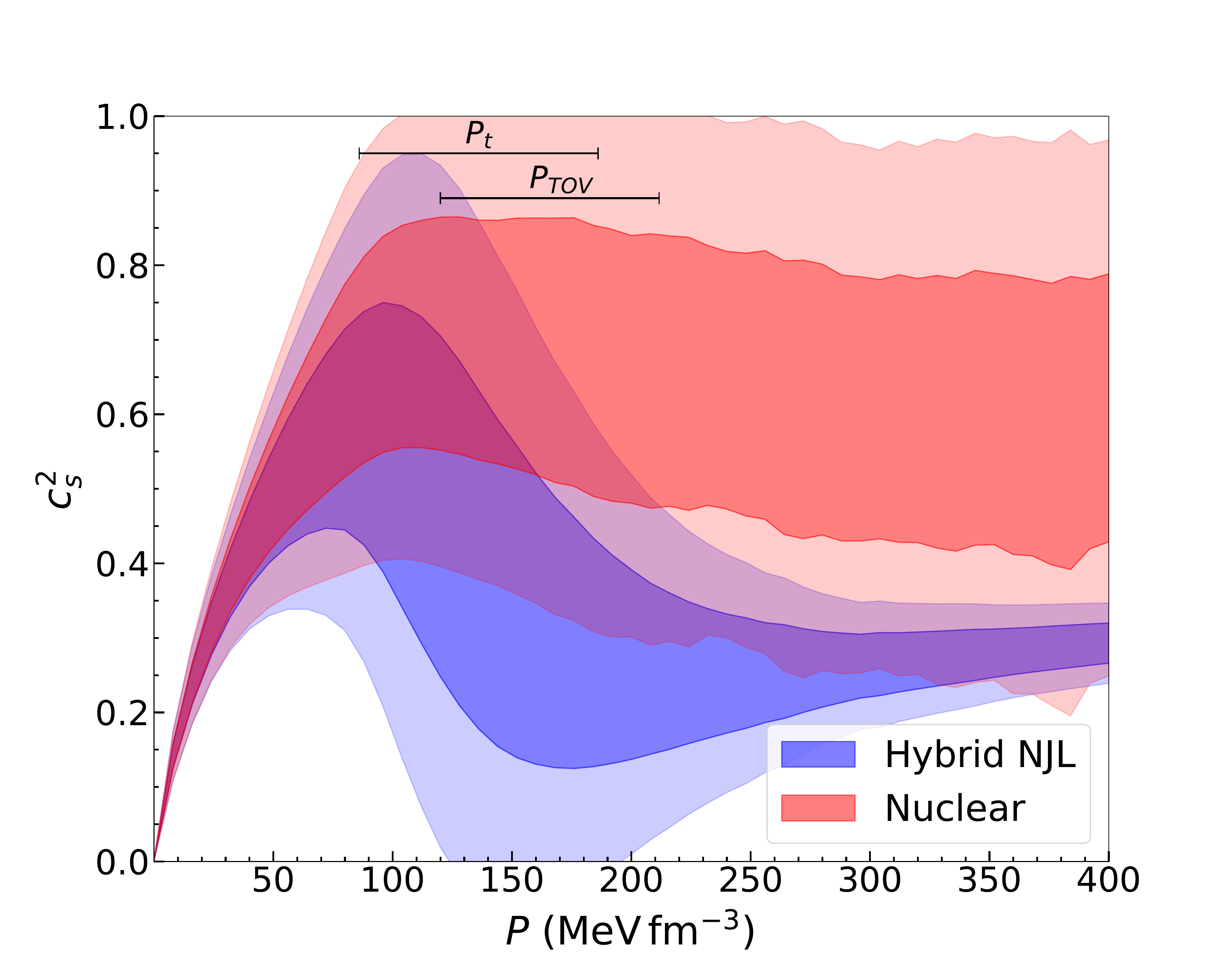}
\caption{Comparison of the posterior sound speed evolution as a function of pressure in the hypothesis of presence (blue) or absence (red) of PT. The darker (resp lighter) regions correspond to 1$\sigma$ (resp 2$\sigma$) uncertainty areas.
The average and 1$\sigma$ uncertainty on the NQ transition pressure $P_{t}$ as well as on the central pressure of the maximum mass configuration $P_{TOV}$ (in the hybrid hypothesis) are also displayed for reference.}
\label{Compcs}
\end{figure}

The resulting posterior distributions of the EoS and the sound speed are shown in Figs. \ref{compEOS} and \ref{Compcs}, comparing the two different hypotheses on the composition of the NS core.
At low densities, the hybrid and nuclear EoSs are in perfect agreement, which is reasonable since in both calculations matter is purely hadronic in this density regime.
The slight preference for stiffer nuclear EoS of the hybrid hypothesis does not bring any meaningful difference on average, as the deviation of the pressure, barely visible in Fig.\ref{compEOS}, is at most of 2.5\%  around ${\rho \approx 1.5\rho_{sat}}$.
However, as we exceed the average PT density (which is about $3.5\rho_{sat}$), the phase transition softens the EoS and lowers the pressure, since the pressure has to stay constant in the whole range of the discontinuity. 

Note that while each individual model presents a pressure plateau as well as a discontinuity in the sound speed, this non analytic behavior is hardly visible in the global distribution, because of the large exploration of both the hadronic and quark parameter space. This underlines the difficulty of getting unequivocal signals of a potential first order phase transition from static observables only.
In particular, the peak structure of the sound speed is due to the fact that 
quark EoSs are usually softer than nuclear ones due to the overall increase in the number of degrees of freedom. 
It is interesting to note that this particular behavior of the sound speed is also a feature of the quarkyonic model \cite{McLerran_2019,margueron_quarkyonic}, a very different approach to describe the deconfinement phase transition at suprasaturation 
densities.

\begin{figure*}[htbp]
\centering
\includegraphics[scale=0.2]{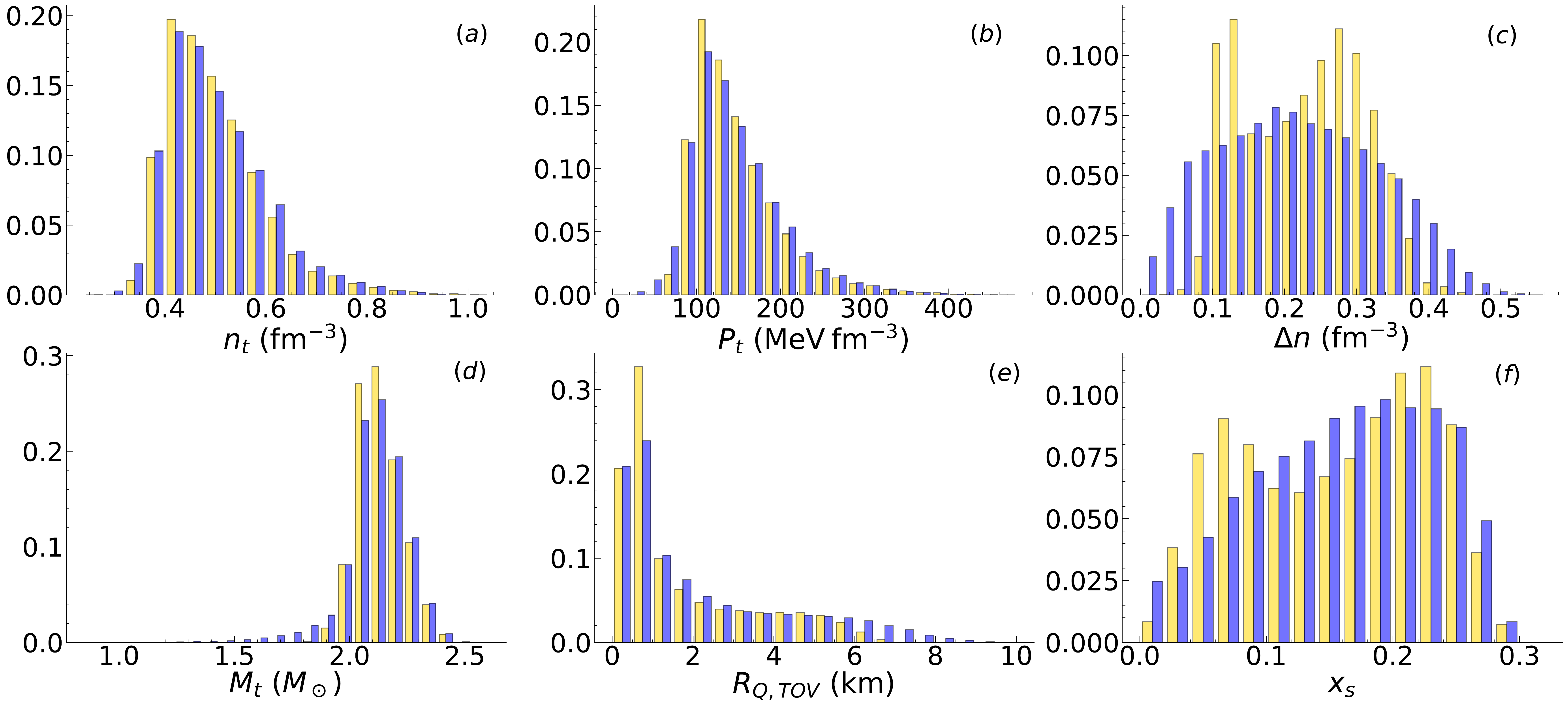}
\caption{Posterior distributions on major properties of hybrid stars, with hypotheses $H_A$ (yellow on the left) and $H_B$ (blue on the right) (see text for details). From left to right and top to bottom: transition density $n_{t}$, transition pressure $P_{t}$, density discontinuity $\Delta n$, total mass at the phase transition $M_t$, radius of the quark core in the maximum mass configuration $R_{Q,TOV}$ and maximal strangeness at the star center $x_s$.  Same normalization as Fig. \ref{quarkparams}.}
\label{trans}
\end{figure*}

\begin{table}[h!]
\centering
\begin{tabular}{|c|c|c|c|c|c|} 
   \hline
    Pulsar &  Model & Mean & $\sigma$ & Min & Max \\
    \hline
    J0030 & Hybrid & 12.9 & 0.4 & 10.6 & 14.3 \\
    \cline{2-6} 
    & Nuclear & 12.9 & 0.4 & 11.1 & 14.7\\
    \cline{2-6} 
    & NICER & 13.1 & 1.2 & 8.9 & 19.7\\
    \hline
    J0740 & Hybrid &12.9 & 0.4 & 10.5 & 14.3  \\
    \cline{2-6} 
    & Nuclear & 13.1 & 0.4 & 11.1 & 14.8\\
    \cline{2-6} 
    & NICER & 13.9 & 2.1 & 2.8 & 26.6\\
    \hline
\end{tabular}
\caption{Mean, standard deviation, and minimum and maximum values of the posterior distributions of the radius (in km) of the J0030 and J0740 pulsars, for our two different models and the analysis of NICER data.}
\label{Nicerradii}
\end{table}

\subsection{NS static properties}

The posterior distributions of various properties of the hybrid stars are displayed in Fig.\ref{trans}, again comparing the results for the two different hypotheses $H_A$ and $H_B$ for the $B^\star$ parameter (see Sec. \ref{sec:Bstar}). We see that the transition density $n_t$ and transition pressure $P_t$ are well peaked around a most favored value of about 0.42~fm$^{-3}$ for $n_t$ and 115~MeV\,fm$^{-3}$ for $P_t$. 
Similar to what is observed on Fig.\ref{quarkparams}, the inclusion of a possible additional bag pressure $B^\star$ has a very marginal effect on these distributions.
We can observe, however, that very early PTs (${n_t\lesssim2n_{sat}}$) cannot be obtained with hypothesis~$H_A$, as they are only made possible by large positive values of $B^\star$.
One has to keep in mind that $B^\star$ would have had a much more important effect on the location of the PT if we had not chosen a prior for $B^\star$ centered around 0.

Concerning the distribution of the density discontinuity $\Delta n$, we observe an interesting double peak structure which gets suppressed once $B^\star$ is allowed to vary.  
This behavior is linked to the effect of the vector couplings on the PT, and will be better understood from the correlation analysis that will be shown later in this section. We also note that in hypothesis $H_B$ both very small {($\Delta n \approx 0$)} and very large {($\Delta n > 0.4$~fm$^{-3}$)} are enabled by the additional freedom, which prefigures an important correlation between this quantity and $B^\star$.

In the bottom panel, we show the distributions of the mass at the transition $M_t$, radius of the quark core $R_{Q,TOV}$ and central strangeness fraction ${x_s = n_s/(n_u+n_d+n_s)}$ in the maximum mass configuration. We observe that, for most EoSs, quarks only appear in very massive stars above 2$M_\odot$. The inclusion of $B^\star$ allows for quark to appear in stars of much lower mass, even below the canonical mass of $1.4M_\odot$ in some cases, but these models remain statistically insignificant. 
We also observe, independently of the hypothesis $H_A$ or $H_B$, that quark cores tend to remain relatively small, even in the maximum mass configuration. This is associated with the fact that the phase transition (and in particular the density discontinuity) considerably softens the EoS and makes quark matter cores quickly unable to balance the gravitational pressure. In addition, because our quark matter EoSs are overall softer than the nuclear ones, stars with large cores may struggle to reach the 2 solar mass threshold and are therefore more often rejected by the associated filter.
However, we remark that it is still possible to build stars above 2 solar masses with quark cores that reach about half of the star's radius. These conclusions might also be relaxed by the presence of a mixed phase smoothening the first-order PT and allowing quarks to appear at lower densities than the ones predicted in our framework \cite{Masuda2013,Maslovpasta,Blaschke2020}. 

On the last panel, we see that it is possible to reach a broad range of strangeness in the quark core, from $x_s=0$ (no strangeness allowed) and sometimes nearly reaching $x_s=1/3$ (flavor-symmetric quark matter). This diversity can be explained by our large exploration of the possible behaviors of the quark~EoS, with the parameter $\xi_\rho$ in particular playing an important role in the flavor balance at high density.
As a consequence, strange quarks pretty much always appear in the heaviest stars in our model, with a maximal strangeness content of about 0.2 on average.
Interestingly, the central strangeness distribution, just like $\Delta n$, exhibits a double peak structure that gets blurred out once the freedom on $B^\star$ is included in the model.

\begin{figure}[h!]
\centering
\includegraphics[scale=0.13]{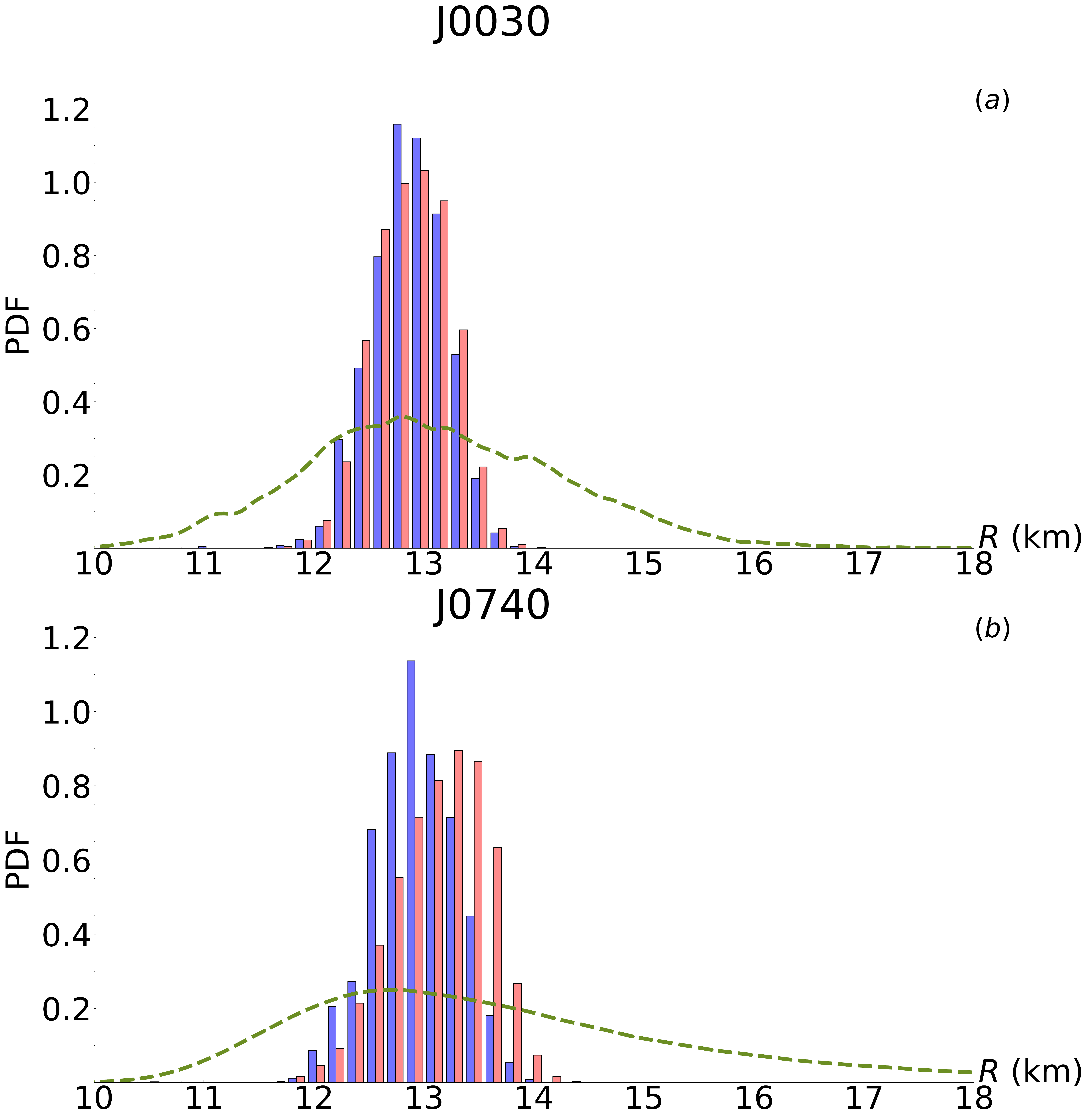}
\caption{Comparison of the posterior distributions of our analysis with both hybrid (blue histogram on the left) and purely nuclear (red histogram on the right) hypotheses and the NICER results of Miller \textit{et al.}~\cite{miller2021} (green dashed line) for the radii of the J0030+0451 ($M=1.44^{+0.15}_{-0.14}M_\odot$) and J0740+6620~($M=2.08^{+0.07}_{-0.07}M_\odot$) pulsars. Each distribution was normalized to unit area.}
\label{Nicercompar}
\end{figure}

In order to compare the results of our analysis with current available data on NS radii, we calculated the radii distributions for two different pulsars: PSR~J0030+0451~{($M=1.44^{+0.15}_{-0.14}M_\odot$)} and PSR~J0740+6620 {($M=2.08^{+0.07}_{-0.07}M_\odot$)},~whose masses have been recently evaluated experimentally using relativistic Shapiro time delay measurements~\cite{Shapiro1,Shapiro2}. 
The radii of these two pulsars have been~estimated via x-ray measurements with the NICER telescope~\cite{NICER,Riley,miller2019,miller2021}.
In Fig.\ref{Nicercompar}, we compare the posterior distributions of radii of each pulsar from the NICER analysis of Miller \textit{et al.}~\cite{miller2019,miller2021} to our results, with and without the introduction of a NQ phase transition.
We first see that the mean values of the distributions are very much in agreement with the NICER measurements for both pulsars and in both of our approaches. However, the distributions inferred from NICER data are much wider than the ones obtained with our analysis, with a standard deviation about 3~times larger for J0030 and up to 5~times larger for J0740  (see Table \ref{Nicerradii} for a summary of the comparison).
Therefore, we conclude that the present experimental uncertainties on radii from x-ray measurements are still too large to put significant additional constraints on the theoretical models, and unfortunately do not bring more information than what can already be inferred from our knowledge of nuclear physics and the analysis of the GW170817 signal.

It is also interesting to compare the results obtained with the two assumptions on the content of the stars. For J0030, there is very little difference noticeable, which is to be expected since the mass of this pulsar is considerably lower than the average mass at which the phase transition occurs (see the bottom left panel of Fig.\ref{trans}). 
Therefore, in both hypotheses we expect J0030 to be made entirely of nuclear matter, whose parametrization will only differ through the quark slight preference for stiffer nuclear EoS (Fig.\ref{NEP}), and eventually does not affect the radius distribution. In contrast, we see a meaningful difference in the posteriors of J0740, with the hybrid models predicting slightly smaller radii than the purely nuclear models. 
This is a direct consequence of the PT, which softens the EoS and reduces the radius as mass increases. However, since most models cannot sustain large quark cores, the difference on the predicted radius remains small.

\begin{figure}[htbp]
\centering
\includegraphics[scale=0.44]{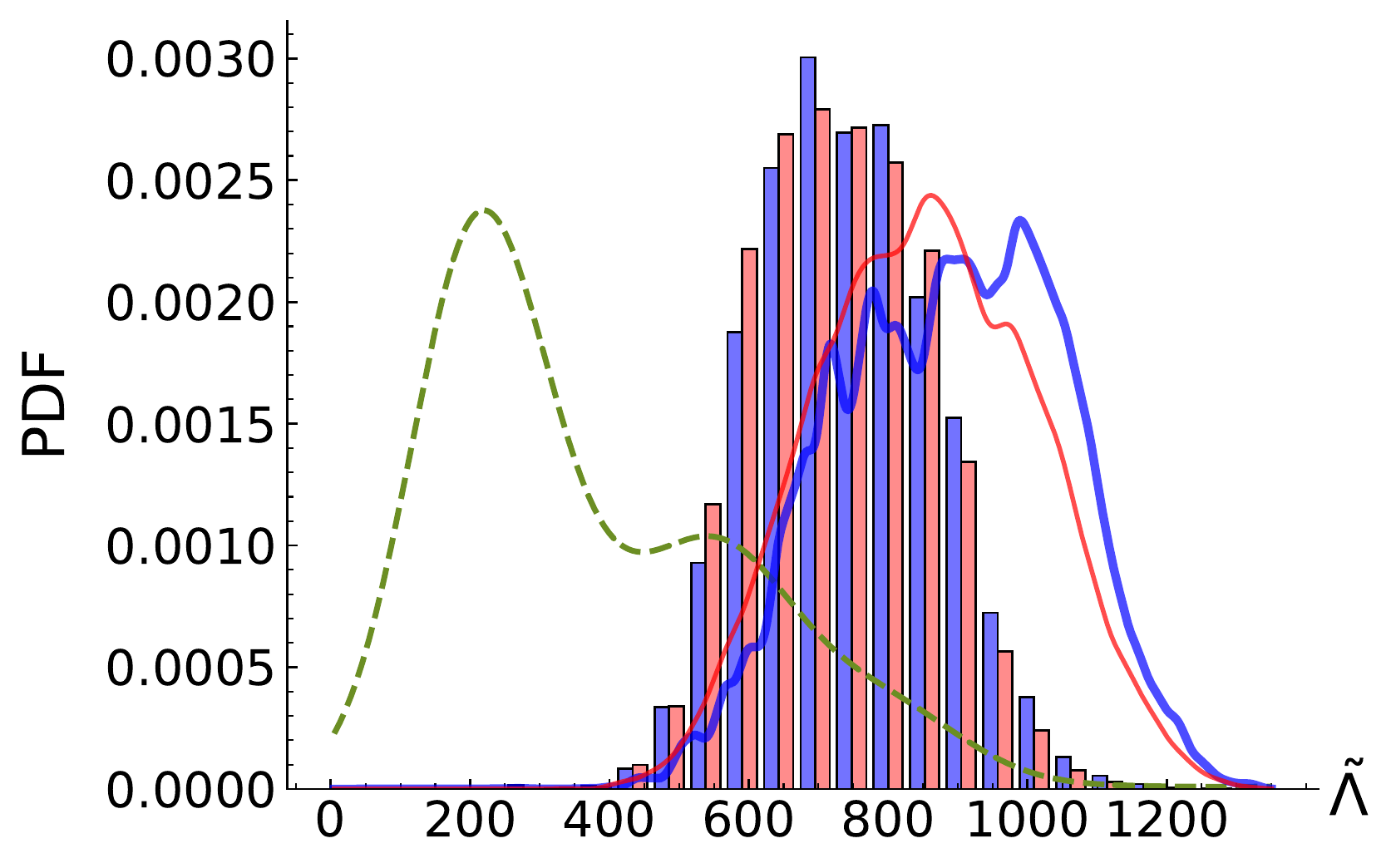}
\caption{Comparison between the GW170817 $\tilde{\Lambda}$ PDF from the LVC analysis of the gravitational wave signal \cite{Abbott19} and our posterior distributions with the hybrid (blue histogram on the left) and purely nuclear (red histogram on the right) hypotheses. The solid lines give the distributions obtained when the LVC filter [weights of Eq.(\ref{GWweight})] is not included. The thick blue line is associated with the hybrid case, and the thin red line with the purely nuclear case. Each distribution was normalized to unit area.} 
\label{170817}
\end{figure}

In Fig.\ref{170817}, we compare the posterior distributions of the weighted-average dimensionless tidal deformability $\tilde{\Lambda}$ from the two hypotheses on the star content, to the posterior distribution of~\cite{Abbott19} extracted by Bayesian inference from the GW170817 signal, given by the green line. This is exactly the probability distribution function appearing in the calculation of our Bayesian weights from GW170817 in Eq.(\ref{GWweight}). In order to make a legitimate comparison, we show for each hypothesis the posterior distributions before (full lines) and after (histograms) the implementation of the GW170817 filter.
Similarly to previous analyses employing a large exploration of the nucleonic EoS parameter space~\cite{Guven2020,Hoa2}, we can see that our prior distribution only agrees with the high $\tilde{\Lambda}$ part of the experimental spectrum; that is it suggests considerably high stiffness within the interval compatible with the GW observation ($\tilde{\Lambda}<800$ at the 90\% level, with a low-spin prior and using the PhenomPNRT waveform model in Ref.\cite{Abbott19}). 
Once the GW170817 filter is applied, our posterior estimation is ${\tilde{\Lambda} = 740 \pm 127}$ in the hybrid hypothesis and ${\tilde{\Lambda} = 728 \pm 124}$ in the purely nuclear case.
Again, including the possibility of a first order phase transition towards quark matter (blue curve and histogram) only marginally modifies the distributions obtained supposing a purely nucleonic content (blue curve and histogram). This is due to the fact that, in the present model, hybrid stars are only realized for very high NS mass which were not explored by the GW170817 event.

\begin{figure}[h!tb]
\centering
\begin{subfigure}
	\centering	
	\includegraphics[scale=0.67]{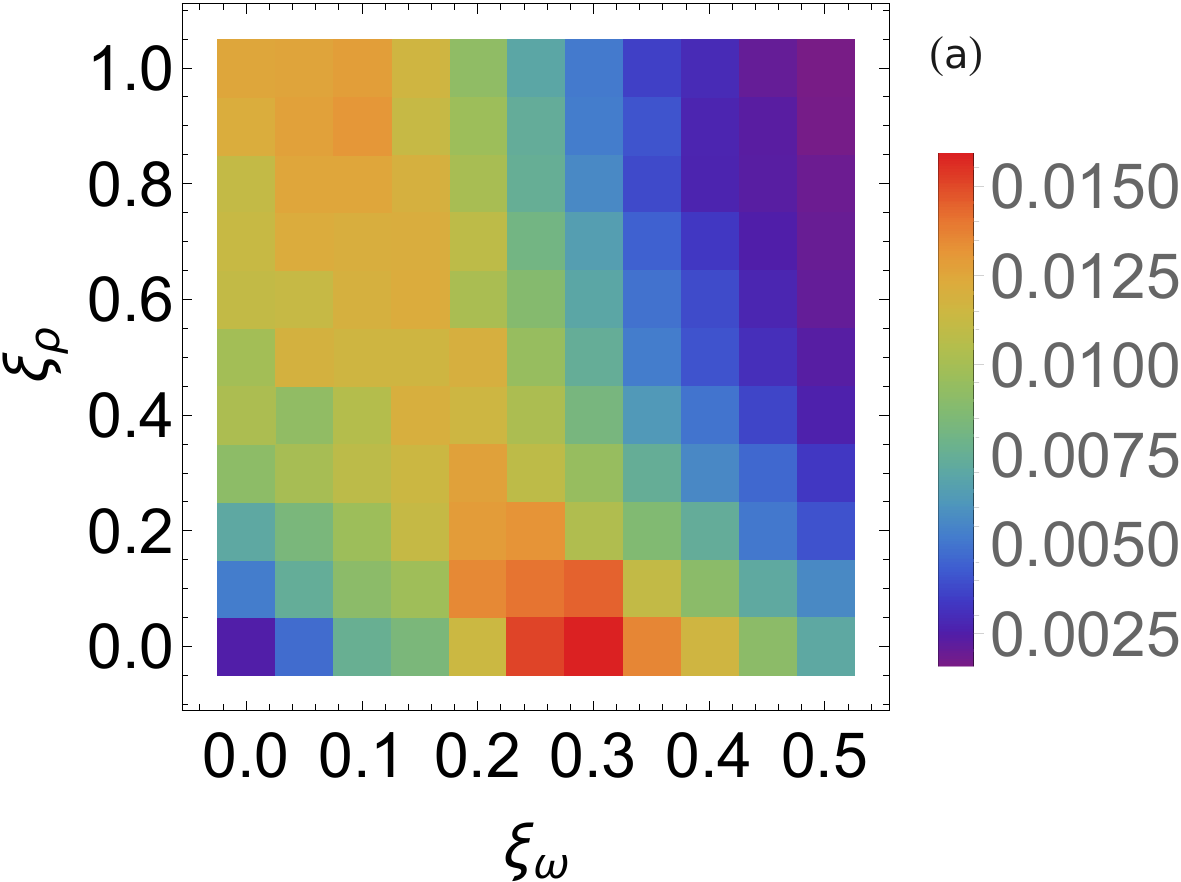}
\end{subfigure}%
\begin{subfigure}
  \centering
  \includegraphics[scale=0.67]{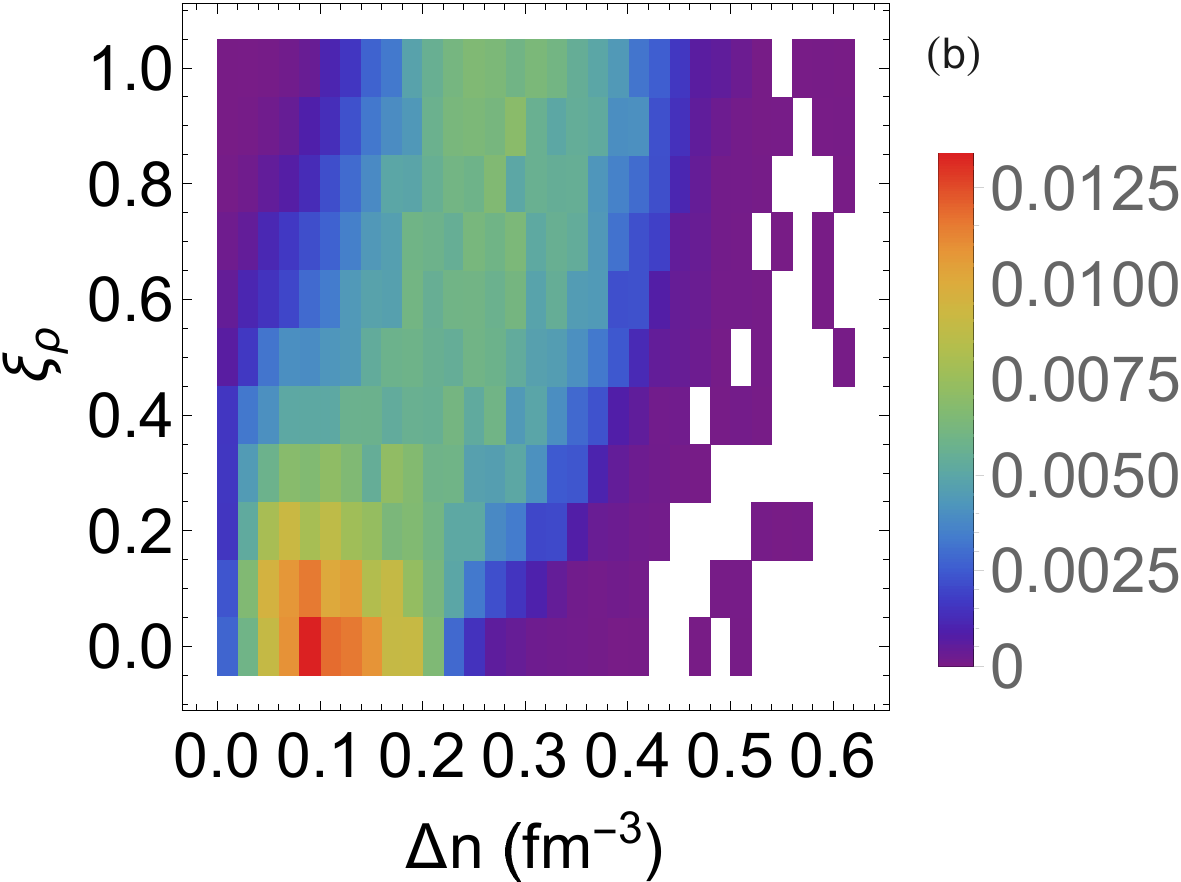}
  
\end{subfigure}
\begin{subfigure}
  \centering
  \includegraphics[scale=0.67]{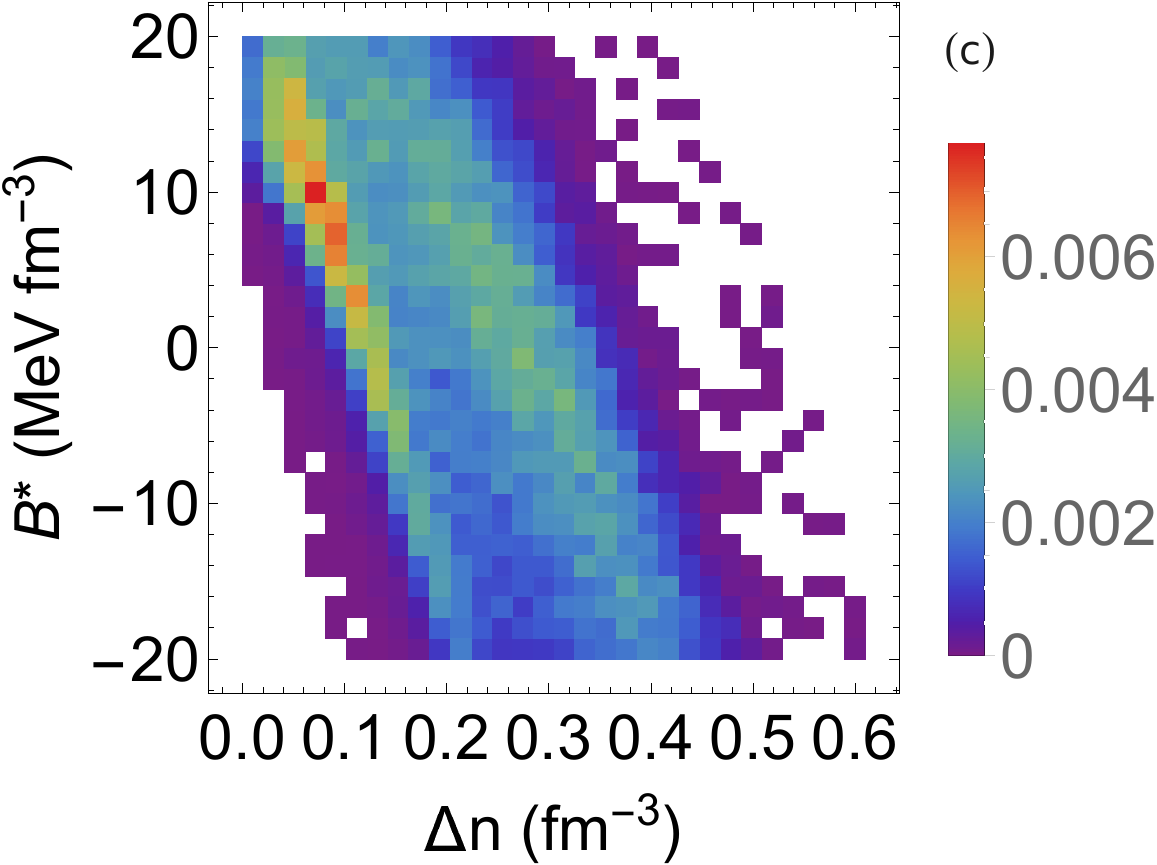}

\end{subfigure}
\caption{2D histograms highlighting the correlation between the quark model parameters and the EoS density discontinuity $\Delta n$. The normalization was chosen such that the sum of each bin height is equal to~1.}
\label{Correlations}
\end{figure}

\vspace{-0.1cm}

\subsection{Parameters and observables correlations}  

We have calculated the Pearson correlation coefficients among all the different model parameters and neutron star observables introduced above, and show in this section only the quantities where a non-negligible correlation was observed. All results shown in this section assume a PT to quark matter with the hypothesis~$H_B$.

In Fig.\ref{Correlations} are shown different two-dimensional(2D) histograms highlighting striking correlations between the parameters. On the top panel, we see that there is a clear anticorrelation between the two different vector channel couplings $\xi_\omega$ and $\xi_\rho$. Both of these parameters tend to stiffen the quark EoS and consequently increase the density of the phase transition. Therefore, if both of them are too low, the PT happens relatively early, but the quark EoS is not stiff enough to reach high $M_{TOV}$. On the other hand, if they are simultaneously high, the quark EoS becomes too stiff and the PT becomes unreachable by compact star densities. The latter effect could be mitigated if we allowed $B^\star$ to take very large values in order to decrease \enquote{manually} the transition density, and allow for stiff quark matter to occur at lower densities. It is also interesting to notice that the canonical values of the vector couplings obtained via Fierz transformations {($\xi_\omega=\xi_\rho=0.5$)} are not favored once we require that quarks must appear inside neutron stars and we apply the astrophysical constraints, highlighting again the importance of the freedom on both of these parameters in our framework.

In the middle and bottom panel we show the dominant correlations between the baryonic density jump $\Delta n$ associated with the phase transition, and the parameters of the quark phase. 
These pictures help to understand the different role played by the vector couplings $\xi_\omega$ and $\xi_\rho$, in particular the role of  $\xi_\omega$ discussed after Fig.\ref{quarkparams}. 
We~can see that, due to the anticorrelation between $\xi_\omega$ and $\xi_\rho$ discussed above, viable hybrid solutions can be obtained even in the case $\xi_\omega=0$, provided the vector-isovector coupling is as important as the scalar coupling. However, because strong interactions in the $\rho$ channel also allow for strange quarks to appear at much lower densities, the quark EoS is significantly  softened at high densities, such that the development of a substantial quark core is hindered.
As observed above and showed by the positive correlation $\xi_\rho-\Delta n$, 
these high couplings push the transition towards very high density values 
that might not be reached in the core of the star. 
As a consequence, only very specific stiff nuclear EoS allowing a sufficiently early phase transition can lead to a non vanishing quark core. 
This specific class of models gives rise to the second peak in the $\Delta n$ and $x_s$ distribution of Fig.\ref{trans}, and to the second branch visible in the correlation ${B^\star-\Delta n}$ (lower panel of Fig.\ref{Correlations}).
Conversely, if the extra stiffness needed in the quark EoS to produce highly massive NS is obtained via the vector-isoscalar coupling ($\xi_\rho=0$ and $\xi_\omega\approx 0.3$), the density jump is relatively small, the extra softness produced by the phase transition negligible, and a much larger class of nuclear EoS are compatible with the possibility of a PT. This category of EoS will also lead to the formation of sizable quark cores in heavy stars. 
Note that when the effective bag parameter is removed, this effect is even more striking, since the additional freedom provided by $B^\star$ tends to wash away the correlations between the parameters. The lower panel of Fig.\ref{Correlations} shows indeed that $B^\star$ is also largely correlated to $\Delta n$ (remember that $\Delta n$ vanishes when $B^\star = B_{max}$).
In particular, positive values of $B^\star$ allow the phase transition to be pushed towards lower densities as seen in Fig.\ref{pressdif}, even for large values of $\xi_\rho$. Consequently, the two branches in the $B^\star-\Delta n$ plane merge together.
This effect explains why the double peaked structure of Fig.\ref{trans} fades away when the effective bag pressure is added to the parameter space.

\begin{figure}[htbp]
\begin{subfigure}
    \centering
    \includegraphics[scale=0.68]{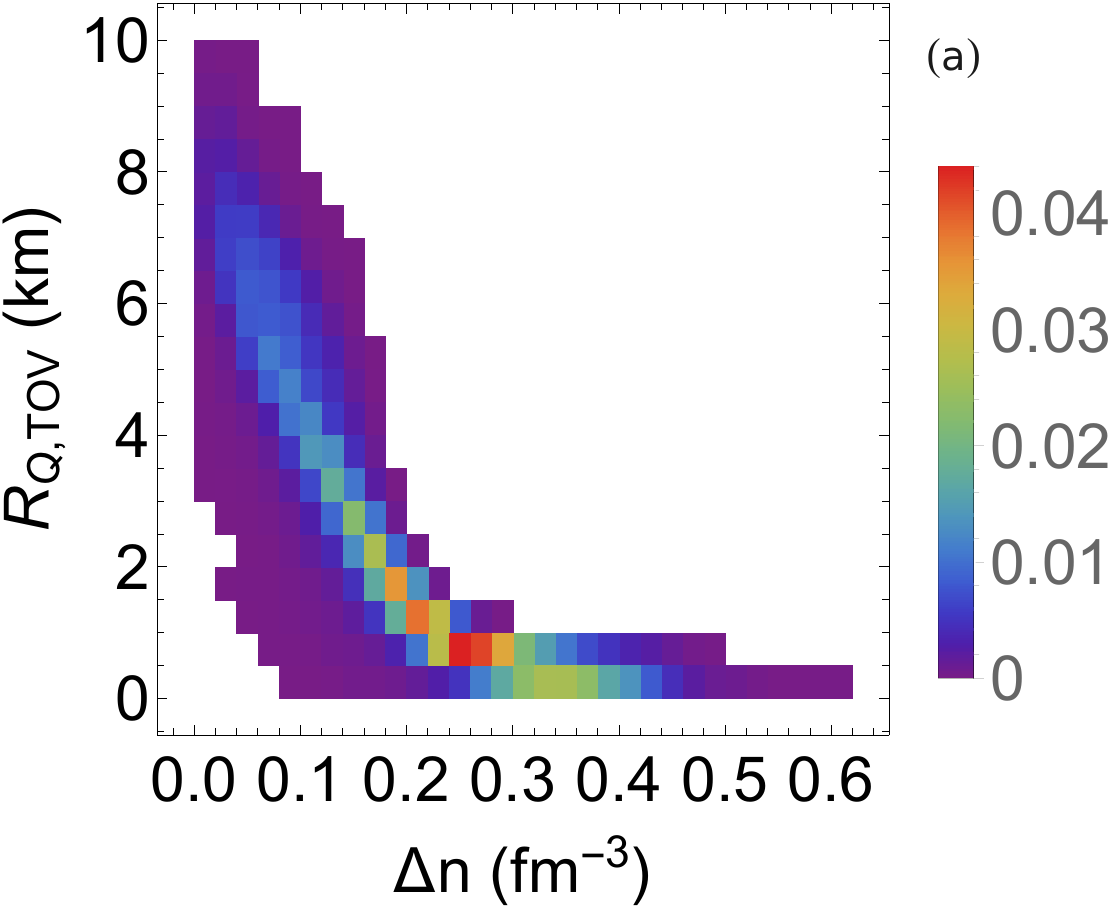}
\end{subfigure}
\begin{subfigure}
    \centering
    \includegraphics[scale=0.68]{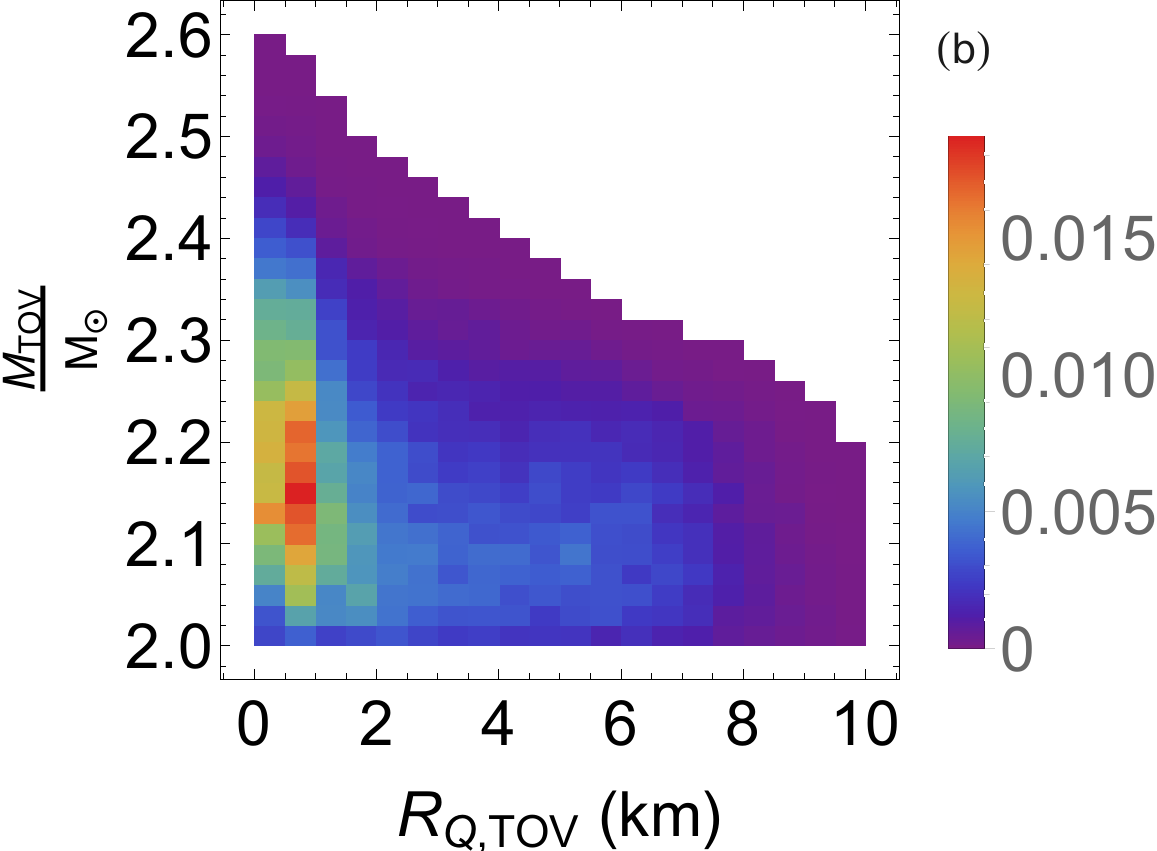}
\end{subfigure}
\caption{2D correlation histograms between the radius of the quark core in the maximum mass configuration $R_{Q,TOV}$ with density discontinuity $\Delta n$ and the maximum mass $M_{TOV}$. Same normalization as Fig. \ref{Correlations}.}
\label{rqmax}
\end{figure}

2D histograms can also reveal interesting properties of hybrid stars as predicted by the NJL model. 
In particular, in Fig.\ref{rqmax} we show that the size of the quark core is strongly correlated to the density discontinuity $\Delta n$ of the phase transition. 
Indeed, we observe that large discontinuities result in a very low stability domain for the quark phase, which consequently cannot reach a substantial proportion of the star. This further amplifies our previous conclusions above on the effect of the vector couplings.
A similar correlation (not shown here because less striking) holds between $R_{Q,TOV}$ and the transition density $n_t$.
This correlation is very easy to understand; if the transition happens at very high densities close to the core density for the maximum mass, there is no room for quarks to appear in the core. Inversely, a PT at low density means that quark matter rapidly takes over as the main constituent of the star, and large quark cores can be formed before reaching gravitational instability. These findings are in good agreement with Ref.\cite{Ferreira2020}. 

\newpage

In addition, we highlight an interesting feature of hybrid stars in the correlation between the size of the quark core and $M_{TOV}$.
In the bottom panel of Fig.\ref{rqmax}, we remark that it is impossible with our model to produce stars that are both very massive {(${M\gtrsim 2.5M_\odot}$)} and holding a sizable quark core.
Therefore, our hypothesis would be in strong disagreement with the results of the analysis of the GW190814 gravitational wave event~\cite{GW190814} if it turned out that the lower mass component {(2.50\,-\,2.67~$M_\odot$)} of the merger was a compact star. This is again in good agreement with the findings of~Ref.\cite{Ferreira2020}.

\begin{figure}[htbp]
\centering
\begin{subfigure}
	\centering	
	\includegraphics[scale=0.7]{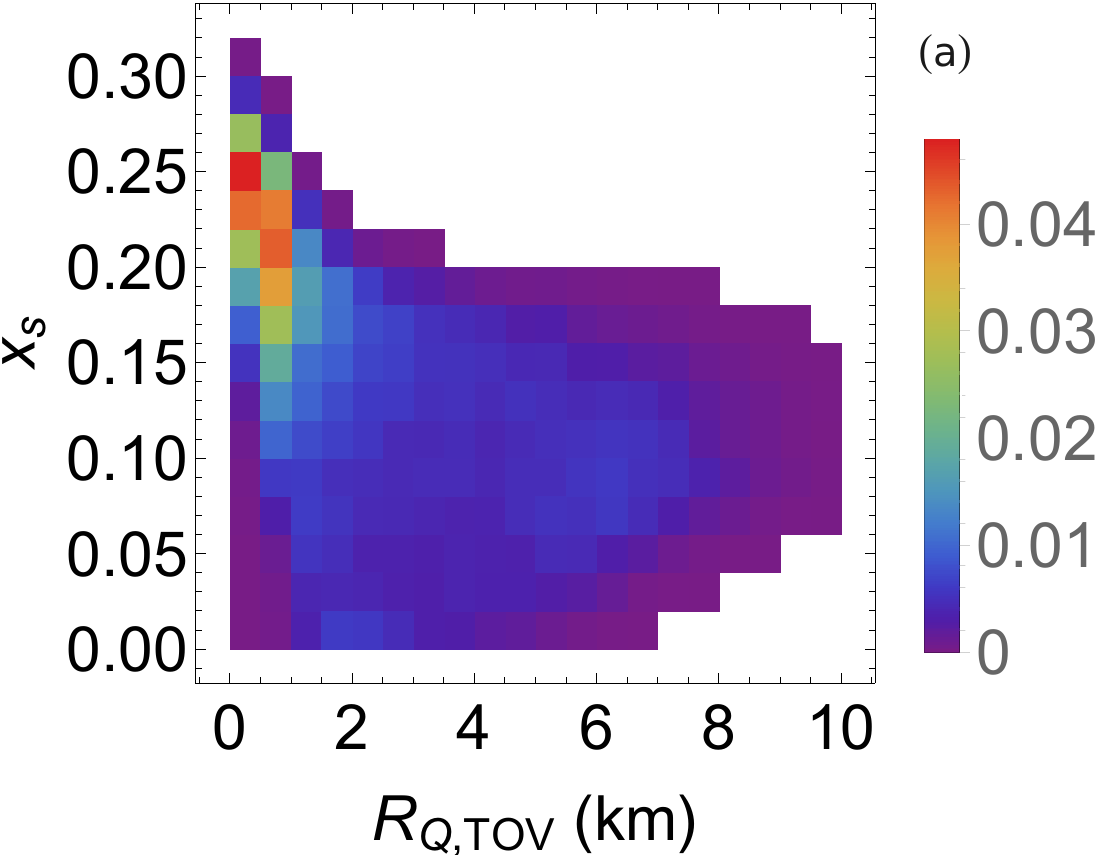}
	
\end{subfigure}%

\begin{subfigure}
  \centering
  \includegraphics[scale=0.7]{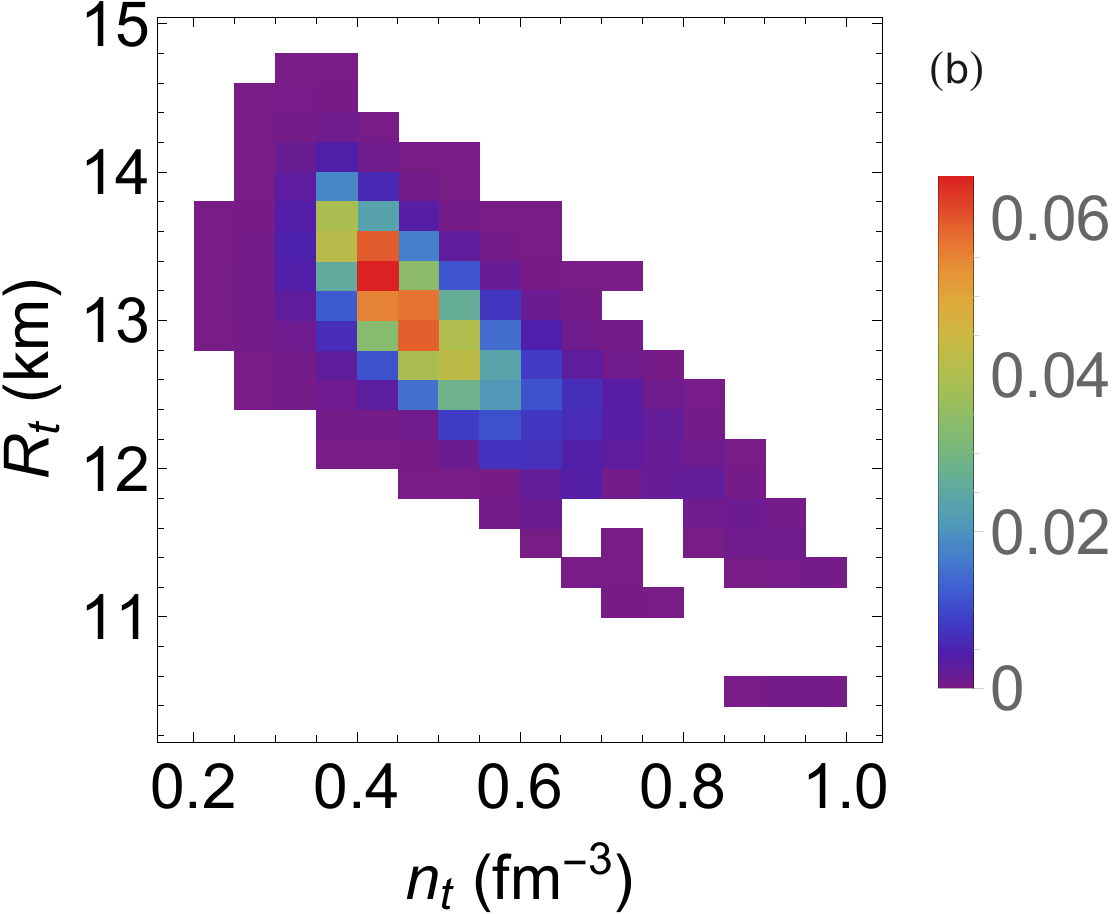}
  
\end{subfigure}
\caption{2D correlation histograms of the radius of the quark core $R_{Q,TOV}$ vs the strangeness fraction at the center of the star $x_s$, as well as transition density $n_t$ vs radius at the transition $R_t$. Same normalization as Fig. \ref{Correlations}.} 
\label{fig:last}
\end{figure}

Finally, Fig.\ref{fig:last} shows the correlation between the size of the quark core and the strangeness fraction for the maximal mass configuration (upper panel), and the correlation between the total star radius at the PT and the transition density (lower panel).
The upper correlation shows that, though our hybrid stars are characterized by an important strangeness content (see also Fig.\ref{trans} above), an important extension of the quark core is only possible for a limited amount of strangeness, since strangeness tends to destabilize the quark core as the EoS becomes too soft. This shows again that large values of $\xi_\rho$, which typically result in a strong PT (\textit{i.e.}, with large $\Delta n$) to a strange-rich quark EoS, are not compatible with large quark cores.

The negative correlation between the radius $R_t$ and the transition density illustrates the somewhat paradoxical feature that while quarks indeed soften the EoS and reduce the radius of the mass-radius sequence, they are more likely to appear at low densities if the nuclear EoS is stiffer (see Fig.\ref{NEP}), that is if the nuclear EoS itself produces large radii.  
This means that measuring a large radii at some mass $\approx 2 M_\odot$ would disfavor the presence of a deconfined core at said mass, but make it more likely to appear at larger masses. This statement is, however, mitigated by the fact that stiff EoSs are known to reach the TOV maximum mass configuration at considerably smaller densities than soft EoSs, such that even small transition densities may not be attainable in NS cores.

\section{Summary and conclusions}

In the present paper, we have performed an extensive Bayesian analysis on the different characteristics of hybrid stars, as obtained if the quark matter is described by the NJL effective model of QCD, and the phase transition is obtained by the Maxwell construction.

A partially agnostic approach is employed for the hadronic part of the EoS, using a flexible meta-modelling technique that allows exploring all the possible density dependences compatible with an analytic behavior of the dense matter energy functional, and at the same time respects the nuclear physics constraints imposed by \textit{ab initio} modeling of homogeneous matter, as well as nuclear mass measurements.

Concerning the quark EoS, the NJL model is built to respect the flavor symmetry constraints of QCD and the parameters are fixed by vacuum meson properties, with the exception of the vector-isoscalar and vector-isovector couplings taken as free parameters.
A varying effective bag constant which allows extra freedom in the localization of the phase transition is also introduced. 

A large variation of the parameter space, both in the hadronic and in the quark sector, produces general predictions for the properties of hybrid stars, with a likelihood conditioned by the constraints given by the different recent observations on static NS properties from radio, X-rays and gravitational waves. 

In agreement with previous studies, we confirm that present observations on the mass, radii, and tidal polarizability are not constraining enough to discriminate between the purely nucleonic scenario and the transition towards quark matter. 

We additionally find that the nuclear matter properties are only slightly modified by the condition that a quark core exists at least in the most massive NS. 
However, an indication towards the existence of hybrid stars could be given by an increased stiffness of the nuclear EoS in the isoscalar sector, with respect to the present values extracted from the giant monopole resonance excitation at subsaturation densities. Such measurements are potentially available through relativistic heavy ion collisions \cite{hades}.  

Our study stresses the importance of vector interactions in quark matter in order for hybrid stars to reach sufficiently high masses. We also report an important correlation between the vector-isoscalar and vector-isovector couplings.

Our prediction for the transition density to quark matter is $n_t=0.48\pm 0.09$~fm$^{-3}$, with a density discontinuity $\Delta n=0.206\pm 0.106$~fm$^{-3}$. This relatively high value of the transition density implies that a quark core would be only present in the most massive NS, ${M_t=2.12\pm 0.15 M_\odot}$. 
For such massive hybrid stars, a non-negligible quark core ${R_Q=2.3\pm 2
.2}$ km is predicted, and the depth of deconfined matter can reach values as large as $\approx 9$ km if the couplings are such that the strangeness content is limited to less than $\approx 10\%$. 

According to our study, the most important signature that distinguishes hybrid stars from nucleonic stars is given by the behavior of the sound speed as a function of the pressure. 
The distribution of the latter is given by a characteristic peaked structure, which is common to other effective approaches dealing with the modeling of deconfinement. 
If this structure will probably be hardly accessible from static observables such as the tidal polarizability, we may expect that the underlying discontinuity will affect in an important way the dynamic properties of the after-merger, that will be accessible by next-generation interferometers.

Finally, it is important to stress that these conclusions are obtained within a specific framework
for the quark EoS. A more important contribution of deconfined matter in compact stars may be obtained
if additional freedom is taken in the quark EoS, for example by the introduction of color superconductivity~\cite{Sedrakian,Alford2007,Blaschke2005}, a crossover treatment of the PT \cite{Blaschke2020,Masuda2013} or other effects of QCD~\cite{Morimoto2020,Alvarez2016,Mattos2021,Mattos2021_bis}.

\section{Acknowledgments}

This publication is part of a project that has received funding from the European Union’s Horizon 2020 research and innovation program under Grant Agreement STRONG – 2020 No 824093. Support by the IN2P3 Master Project MAC is acknowledged. We are grateful to H.Dinh Thi from LPC Caen for providing codes and analysis tools for the nucleonic EoS.

\appendix

\section{Summary of the results}
The posterior average, standard deviation, and minimum and maximum values of the different parameters and observables discussed in the text are summarized in Table~\ref{Meanvalues}.

\begin{table}[htbp]
\centering
\begin{tabular}{|c|c|c|c|c|c|}
\hline
  & Unit & Mean & $\sigma$ & Min & Max  \\
\hline
$n_{sat}$ & fm$^{-3}$ & 0.163 & 0.005 & 0.15 & 0.17\\
\hline
$E_{sat}$ & MeV & -15.90 & 0.55 & -17.18 & -14.50\\
\hline
$K_{sat}$ & MeV & 263.7 &26.6 & 190.0 & 300.0\\
\hline
$Q_{sat}$ & MeV & 184 & 328 & -777 & 1000\\
\hline
$Z_{sat}$ & MeV & 612 & 2004 & -3992 & 5000\\
\hline
$E_{sym}$ & MeV & 30.9 & 1.2 & 27.2 & 34.5\\
\hline
$L_{sym}$ & MeV & 47.4 & 9.2 & 22.7 & 75.3\\
\hline
$K_{sym}$ & MeV & -61 & 94 & -310 & 300 \\
\hline
$Q_{sym}$ & MeV & 1220 & 792 & -958 & 4964\\
\hline
$Z_{sym}$ & MeV & -14 & 2790 & -4995 & 5000\\
\hline
$m^\star_{sat}/m$ &  & 0.702 & 0.059 & 0.6 & 0.8\\
\hline
$\Delta m^\star_{sat}/m$ &  & 0.051 & 0.088 & -0.1 & 0.2\\
\hline
$b$ &  & 4.23 & 2.56 & 1 & 10\\
\hline
$\xi_\omega$ &  & 0.20 & 0.14 & 0 & 0.5\\
\hline
$\xi_\rho$ &  & 0.47 & 0.32 & 0 & 1\\
\hline
$B^\star$ & MeV\,fm$^{-3}$& 1.4 & 10.9 & -20 & 20\\
\hline
$\Delta n$ & fm$^{-3}$ & 0.206 & 0.106 & 0.001 & 0.610\\
\hline
$\Delta \rho$ & $10^{15}$g\,cm$^{-3}$ & 0.50 & 0.27 & 0.001 & 1.7\\
\hline
$n_t$ & fm$^{-3}$ & 0.48 & 0.09 & 0.24 & 0.98 \\
\hline
$P_t$ & MeV\,fm$^{-3}$ & 138 & 50 &  17 & 492 \\
\hline
$\mu_t$ & MeV & 1337 & 86 &  1042 & 1770 \\
\hline
$\rho_t$ & $10^{15}$g\,cm$^{-3}$ & 0.91 & 0.21 & 0.41 & 2.2 \\
\hline
$\mu_{e,N}$ & MeV & 223.2 & 49.9 & 22.8 & 372.3 \\
\hline
$\mu_{e,Q}$ & MeV & 81.9 & 14.7 & 28.5 & 110.1 \\
\hline
$M_{t}$ & $M_\odot$  & 2.12 & 0.15 & 0.87 & 2.59 \\
\hline
$R_t$ & km  & 13.0 & 0.5 & 10.4 & 14.6\\
\hline
$M_{TOV}$& $M_\odot$ & 2.16 & 0.10 & 2.00 & 2.59\\
\hline
$M_{TOV}-M_t$& $M_\odot$ & 0.04 & 0.10 & 0.00 & 1.14\\
\hline
$n_{TOV}$& fm$^{-3}$ & 0.77 & 0.12 & 0.48 & 1.51 \\
\hline
$P_{TOV}$& MeV\,fm$^{-3}$ & 166.2 & 45.9 & 58.1 & 492.8 \\
\hline
$\rho_{TOV}$ & $10^{15}$g\,cm$^{-3}$ & 1.59 & 0.29 & 0.93 & 3.65 \\
\hline
$M_{Q,TOV}$ & $M_\odot$ & 0.13 & 0.23 & 0.00 & 1.62 \\
\hline
$R_{Q,TOV}$ & km & 2.3 & 2.2 & 0.0 & 9.9\\
\hline
$x_{s,TOV}$ &  & 0.15 & 0.07 & 0.00 & 0.31\\
\hline
$R_{1.4}$ & km & 12.9 & 0.3 & 11.1 & 14.0\\
\hline
$\Lambda_{1.4}$ &  & 648 & 113 & 223 & 1164\\
\hline
$R_{2.0}$ & km & 13.1 & 0.4 & 10.6 & 14.5\\
\hline
$\Lambda_{2.0}$ &  & 73 & 19 & 12 & 166\\
\hline
$R_{TOV}$ & km & 12.9 & 0.5 & 10.4 & 14.6\\
\hline
$\Lambda_{TOV}$ &  & 38 & 15 & 9 & 162\\
\hline
\end{tabular}
\caption{Summary of the posterior mean, standard deviation, and minimal and maximal values for the model parameters and various properties of the EoS in the hybrid case with hypothesis $H_B$. For the definitions of the different quantities, see the text. }
\label{Meanvalues}
\end{table}

\newpage

\bibliography{biblio}

\end{document}